\documentclass[preprint,12pt]{elsarticle}
\usepackage{amssymb}
\usepackage{amsmath}
\usepackage{graphicx}
\usepackage{subcaption}
\usepackage{lipsum}
\usepackage[dvipsnames]{xcolor}
\usepackage{soul}
\usepackage{float}
\usepackage{hyperref}
\usepackage{multirow}
\usepackage{lineno}
\usepackage{xurl}
\usepackage{array}
\hypersetup{
    colorlinks=true,
    linkcolor=blue,
    citecolor=blue,
    urlcolor=blue
}
\biboptions{sort&compress} % Add this line for compressed citations
\begin{document}
\begin{frontmatter}

\title{Discerning and quantifying high frequency activities in EEG under normal and epileptic conditions}

\author[a1]{Jyotiraj Nath} %% Author name

%% Author affiliation
\affiliation[a1]{addressline={SVNIT Surat}, city={Surat},
            postcode={395007}, 
            state={Gujrat},
            country={India}}

\author[a2]{Shreya Banerjee} %% Author name
%% Author affiliation
\affiliation[a2]{ addressline={Center for Quantum Science and Technology,\\ Siksha 'O' Anusandhan University},     city={Bhubaneswar},
            postcode={751030}, 
            state={Odisha},
            country={India}}

\author[a3]{Bhaswati Singha Deo} %% Author name
%% Author affiliation
\affiliation[a3]{ addressline={Indian Institute of Technology Kanpur}, 
            city={Kanpur},
            postcode={208 016}, 
            state={Uttar Pradesh},
            country={India}}

\author[a4]{Mayukha Pal} %% Author name%% Author affiliation
\affiliation[a4]{ addressline={ABB Ability Innovation Centre, Asea Brown Boveri Company}, 
            city={Hyderabad},
            postcode={500084}, 
            state={Telengana},
            country={India}}

\author[a2,a5]{Prasanta K. Panigrahi 
\corref{mycorrespondingauthor}}
\ead{pprasanta@iiserkol.ac.in}
\cortext[mycorrespondingauthor]{Corresponding author}
%% Author name%% Author affiliation
\affiliation[a5]{ addressline={Department of Physcial Sciences, Indian Institute of Science Education and Research Kolkata}, 
            city={Mohanpur},
            postcode={741246}, 
            state={West Bengal},
            country={India}}

%% Abstract
\begin{abstract}
We investigate the nature of the modifications in the temporal dynamics manifested in the high-frequency EEG spectra of the normal human brain in comparison to the diseased brain undergoing epilepsy. For this purpose, the Fourier reconstruction is efficaciously made use of after Welch's transform, which helped identify the relevant frequency components undergoing significant changes in the case of epilepsy. The temporal dynamics involved in the EEG signals and their associated variations showed a well-structured periodic pattern characterized by bi-stability and significant quantifiable structural changes during epileptic episodes. In particular, we demonstrate and quantify the precise differences in the high-frequency gamma band (40-100 Hz) present in EEG recordings from neurologically normal participants compared to those with epilepsy. The periodic modulations at two dominant frequencies around 50 Hz and 76 Hz in power spectral density are isolated from high-frequency noise through the use of Welch’s transform, pinpointing their collective behaviors through a phase-space approach. The reconstructed signals from these restricted frequency domains revealed oscillatory motions showing bistability and bifurcations with distinct differences between normal and seizure conditions. These differences in the phase space images, when analyzed through linear regression and SVM-based machine learning models, support a classification accuracy of around 94–95\% between healthy and ictal states using a publicly available EEG dataset from the University of Bonn (Germany). The partial reconstruction of the dynamics as compared to the earlier studies of the full phase space accurately pinpointed the destabilization of the collective high-frequency synchronous behavior and their precise differences in the normal and diseased conditions, avoiding the other chaotic components of the EEG signals.

\end{abstract}

%% Keywords

\end{frontmatter}
\section{Introduction}\label{intro}

%----To remove plagiarism----
Epilepsy is a persistent neurological disorder marked by recurrent seizures affecting more than 50 million individuals worldwide, causing substantial detriments to quality of life \cite{WHO2023, MALAKOUTI2025134}. 
%--------------
% Epilepsy is a chronic neurological disorder characterized by recurrent, unprovoked seizures resulting from abnormal neuronal activity. It affects over 50 million people worldwide and significantly impacts quality of life due to its unpredictable nature and associated cognitive and physical impairments \cite{WHO2023}.
%------------------
The brain's electrical activity, orchestrated by billions of interconnected neurons, plays a vital role in cognitive functions, behavior regulation, and physiological coordination. Disruptions in this intricate system, especially abnormal synchronizations, often manifest as epileptic seizures. Electroencephalography (EEG) is a non-invasive technique that records electrical activity from the scalp and is widely used in clinical and research settings for seizure diagnosis and brain state monitoring. EEG's millisecond-scale temporal resolution enables real-time detection of pathological events, such as epileptiform discharges or rhythmic slowing \cite{Schomer2018}. However, due to its nonlinear and non-stationary nature, the EEG signal poses significant challenges in feature extraction and interpretation. Artifacts, inter-subject variability, and the subtlety of pre-ictal changes make manual analysis labor-intensive and susceptible to subjectivity, thereby motivating the development of automated, data-driven methods for seizure detection and classification.\\Among the canonical EEG frequency bands—Delta (0.5--4 Hz), Theta (4--8 Hz), Alpha (8--13 Hz), Beta (13--30 Hz)—gamma oscillations (30--100 Hz) have emerged as particularly relevant in the context of epilepsy. Gamma-band activity has been linked to both high-level cognitive processes and pathological conditions, with studies reporting increased synchrony or power modulation in this band during ictal periods \cite{Moezzi2022, Tsiouris2021}. These oscillations often exhibit coupled oscillator dynamics, and their temporal patterns serve as potential markers for seizure onset zones \cite{9714339}. Yet, the accurate extraction of such patterns is hindered by the presence of high-frequency noise and the complex superposition of linear and nonlinear dynamics.\\Several studies \cite{9754583,SHEN2022103820,10.3389/fnins.2024.1436619,PAL2022126516} have sought to characterize these dynamics using spectral and nonlinear techniques. Classical methods based on Fourier and wavelet transforms have been employed to isolate frequency-specific anomalies, but they often fall short in capturing transient, multiscale features associated with seizures. To address this, nonlinear dynamic frameworks such as phase-space reconstruction, Lyapunov exponent estimation, and fractal dimension analysis have been introduced to identify the chaotic and bifurcating behavior of epileptic brain states \cite{kannathal2005entropies,rosso2001wavelet}. These techniques provide insight into the complexity of seizure evolution but can become computationally intensive or sensitive to embedding parameters.\\In parallel, machine learning and deep learning models have demonstrated significant promise in automated seizure classification. Traditional approaches rely on handcrafted features followed by classifiers such as support vector machines (SVMs) or random forests \cite{SHEN2022103820,Farawn}. More recent efforts incorporate convolutional neural networks (CNNs) \cite{Duan2022An, elshekhidris} and transformer-based models \cite{9754583, s24113360} to extract spatial and temporal features directly from raw EEG data. One such architecture, GlepNet \cite{10541111}, combines convolutional filters and multi-head attention mechanisms to learn robust global-local representations and has achieved state-of-the-art performance on multiple EEG benchmarks. Despite these advancements, end-to-end models often function as black boxes and require large labeled datasets, limiting their clinical adoption and interpretability.\\Traditional time-frequency analysis methods, such as short-time Fourier transforms and classical periodograms, suffer from poor resolution and high variance in the high-frequency gamma range. These shortcomings lead to spectral leakage, making it difficult to reliably detect gamma-band abnormalities \cite{Jia2020}. Welch's method, a refined periodogram approach, addresses these issues by segmenting signals into overlapping windows and averaging the resulting spectra, thereby enhancing both frequency resolution and noise robustness \cite{Moca2021}. To further characterize the underlying EEG dynamics, nonlinear phase-space reconstruction has been proposed as a promising tool to visualize and analyze the system's evolution in terms of attractor structures and state transitions. Applied to gamma-band filtered signals, this approach can reveal abrupt changes in neuronal synchrony associated with ictal and inter-ictal brain states \cite{Li2021}. While these signal processing techniques provide rich representations, their patterns are often too complex for reliable manual interpretation. This has motivated the use of machine learning for automated classification. Recent in deep learning, particularly Convolutional Neural Networks (CNNs), have demonstrated high performance in EEG-based seizure detection tasks \cite{Roy2022, Ahmad2023}. By encoding EEG dynamics into image-like formats, such as spectrograms or phase portraits, CNNs can learn rich hierarchical features without requiring hand-crafted preprocessing. This makes them especially suitable for classifying nonlinear, high-frequency EEG representations.

The aim of this study is to identify and quantify gamma-band EEG differences between healthy and epileptic conditions, including preictal, interictal, and ictal phases. Key challenges include (i) isolating gamma activity from noisy EEG recordings, (ii) modeling its nonlinear temporal dynamics, and (iii) building a robust classifier for automated diagnosis. Conventional diagnostic pipelines often rely on limited visual inspection or low-dimensional statistical measures, which may miss subtle but critical dynamical patterns. Therefore, we propose an integrated computational framework that combines signal processing, dynamical system reconstruction, and machine learning for improved interpretability and classification accuracy.

This study proposes a novel hybrid approach for EEG-based epilepsy classification. The key contributions are:

\begin{enumerate}
\item Application of Welch's power spectral density estimation to robustly isolate gamma-band oscillations from EEG signals, mitigating the impact of spectral leakage and noise.
\item Reconstruction of phase-space portraits from gamma-band filtered signals to capture nonlinear dynamical features of epileptic and non-epileptic brain states.
\item Transformation of phase-space trajectories into grayscale images suitable for input into an ML Classifier for automatic feature extraction and classification.
\item Comparative analysis using logistic regression and Support Vector Machines (SVMs) to assess classifier robustness across EEG condition pairs.
\item A physically interpretable and computationally efficient framework for distinguishing epilepsy-related EEG patterns using minimal manual preprocessing.
\end{enumerate}

The paper is structured as follows: Section \ref{section2} presents the dataset, preprocessing steps, and proposed methodology. Section \ref{sec:result} discusses the results of the proposed method. Finally, Section \ref{sec:conclusion} concludes the paper and suggests future research directions.

\section{Proposed Method} \label{section2}
\begin{figure}[h!]
\centering
    \includegraphics[scale=0.04]{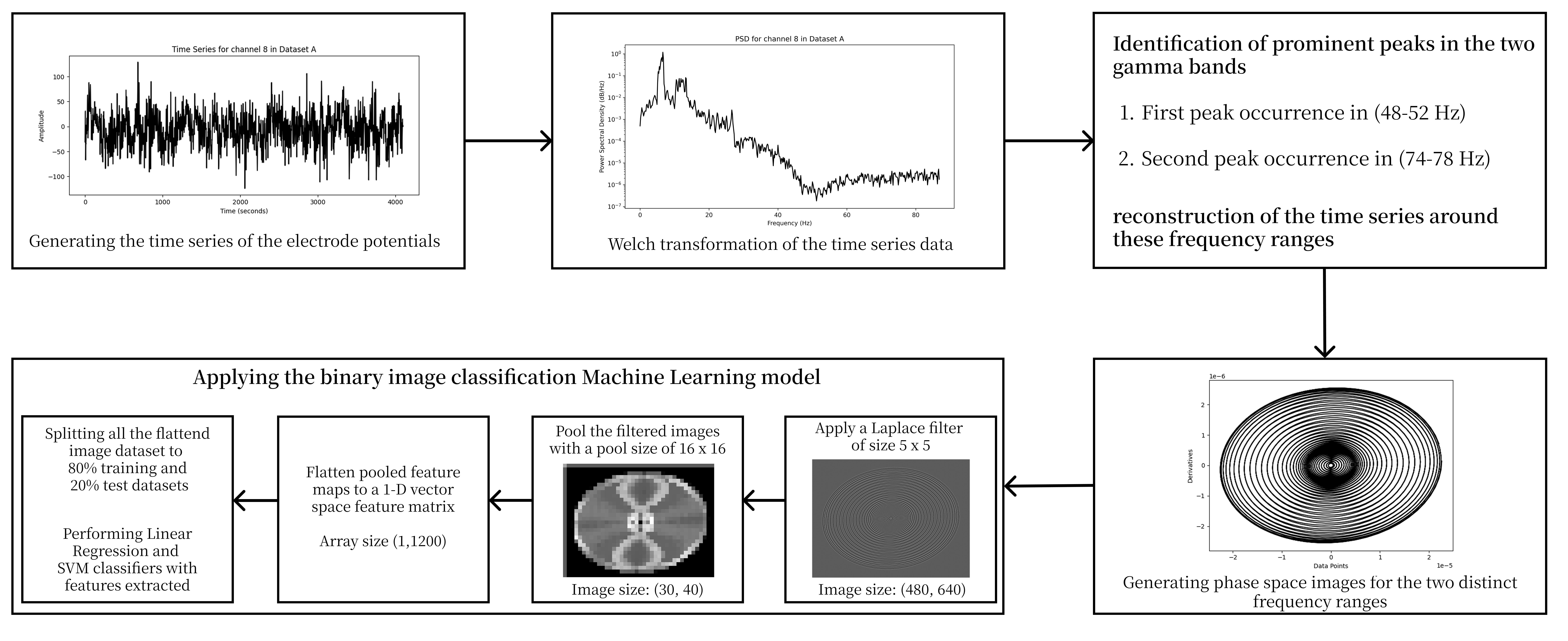}
    \caption{Proposed EEG epilepsy classification workflow: gamma-band identification and time-series reconstruction, phase space imaging, feature extraction, and final SVM/Logistic Regression classification}
    \label{fig:model-workflow}
\end{figure}

An overview of the proposed methodology is presented in Figure \ref{fig:model-workflow}. The approach aims to classify EEG signals into healthy and epileptic categories using a combination of phase space image representations and machine learning classifiers. The EEG data used for analysis is sourced from the University of Bonn dataset~\cite{PhysRevE.64.061907}. The pipeline consists of multiple stages, including spectral analysis of the EEG time series, identification and reconstruction of gamma-band activity, phase space image generation, convolution-based feature extraction, and final classification using Logistic Regression and Support Vector Machine (SVM) models. Each component of the framework is described in detail in the following subsections.
\label{sec:method}
\subsection{Dataset description}
\label{subsec:dataset}
The data used for all the analyses is retrieved from the Clinic and Polyclinic of Epileptology, University Hospital of Bonn, Germany\cite{PhysRevE.64.061907}, collected using a standardized electrode placement scheme (10-20 electrode system)\cite{article_jasper}. The data set consists of individual sets (A, B, C, D, and E) of at least five healthy and five epileptic patients. Each set has 100 single-channel time series segments captured at 173.61 Hz for 23.6 seconds of the time window frame. Extra-cranial data sets A and B are from healthy individuals human with eyes open and closed condition, respectively. Intra-cranial C and D sets are taken from the Hippocampal and Epileptogenic brain regions of epileptic patients. Set E is of patients experiencing a seizure (ictal) condition.

\subsection{Preprocessing}
The raw EEG signals were standardized to mitigate the influence of external confounding factors such as demographic variability, environmental noise, and non-biological artifacts. This involved normalizing each single-channel time series to zero mean and unit variance (z-score normalization), ensuring robust downstream analysis \cite{doi:https://doi.org/10.1002/9780470511923.ch2}. Standardization is critical for EEG preprocessing, as it minimizes inter-subject variability and enhances the generalizability of machine learning models. This approach aligns with established practices in EEG-based epilepsy detection, where amplitude normalization reduces the impact of non-pathological signal variations\cite{https://doi.org/10.1111/j.0013-9580.2005.66104.x}.

% For visualization purposes, each dataset is transformed from a time domain to a frequency domain using the direct Fast Fourier Transform (FFT) method. 
\subsection{Power Spectral Analysis with Welch's Method}
The power spectral density (PSD) quantifies the distribution of signal power across frequency components, enabling identification of pathological oscillatory patterns in EEG \cite{widmann2015digital}. For a discretized EEG signal \( x[n] \) of length \( N \), the classical periodogram estimates PSD as:

\begin{equation}
    P_{\text{periodogram}}(f) = \frac{1}{N} \left| \sum_{n=0}^{N-1} x[n]e^{-j2\pi fn} \right|^2
\end{equation}

where units are $\mu$V\textsuperscript{2}/Hz for voltage-based signals \cite{oppenheim1999discrete}. This estimator suffers from high variance (\( \mathcal{O}(1) \)) and spectral leakage due to finite sampling \cite{welch1967use}. Welch's method addresses these limitations by segmenting \( x[n] \) into overlapping 4-second Hann windows (50\% overlap), applying window functions to reduce leakage, and averaging modified periodograms:

\begin{equation}
    P_{\text{Welch}}(f) = \frac{1}{K} \sum_{k=1}^K \frac{1}{U} \left| \sum_{n=0}^{L-1} w[n]x_k[n]e^{-j2\pi fn} \right|^2
\end{equation}

where \( w[n] \) is the window function and \( U \) the normalization factor. The resulting frequency resolution was \( \Delta f = 0.17\) Hz, with oscillatory power quantified in standard bands: delta (0.5-4 Hz), theta (4-8 Hz), alpha (8-13 Hz), beta (13-30 Hz), and gamma (30-100 Hz) \cite{tenke2015digital}. Gamma-band peaks identified in Welch PSD plots prompted time-domain reconstruction through inverse FFT for phase-space analysis \cite{stam2005nonlinear}.

\subsection{Phase Space Analysis}

Phase Space technique was applied to the EEG time series to investigate the underlying nonlinear dynamics of the brain activity. This enables the unfolding of a system's complete dynamical structure, from a single measurement variable \cite{takens1981detecting, kantz2004nonlinear}. The resulting trajectory in phase space provides a geometric representation of the system's evolution, whereas the shape reveals the nature of the dynamics (e.g., periodic, chaotic, or stochastic).

Phase Space analysis for this study was specifically conducted on the gamma-band (30-100 Hz) component, reconstructed from Welch PSD peaks. A two-dimensional phase space portrait was constructed for each signal. The portrait is formed by plotting the signal's instantaneous amplitude, $x[n]$, against its first time derivative, $\dot{x}[n]$. The time derivative was numerically estimated using a second-order central difference formula:
\begin{equation}
    \dot{x}[n] = \frac{x[n+1] - x[n-1]}{2\Delta t}
    \label{eq:central_difference}
\end{equation}
where $\Delta t$ represents the sampling interval of the discrete signal $x[n]$. This $(x, \dot{x})$ plane serves as a projection of the system's high-dimensional state space.

In the final pipeline step, the dynamical information contained within the phase space trajectories were converted into a format suitable for machine learning analysis. Each reconstructed trajectory was systematically rendered into a two-dimensional grayscale image with a resolution of $512 \times 512$ pixels. These images, which encapsulate the geometric features of the EEG dynamics, were then used as inputs for the subsequent classification using a Machine Learning classifier.

\subsection{Classification} \label{Classification}

\begin{figure}[!hbt]
    \centering
    \includegraphics[scale=0.04]{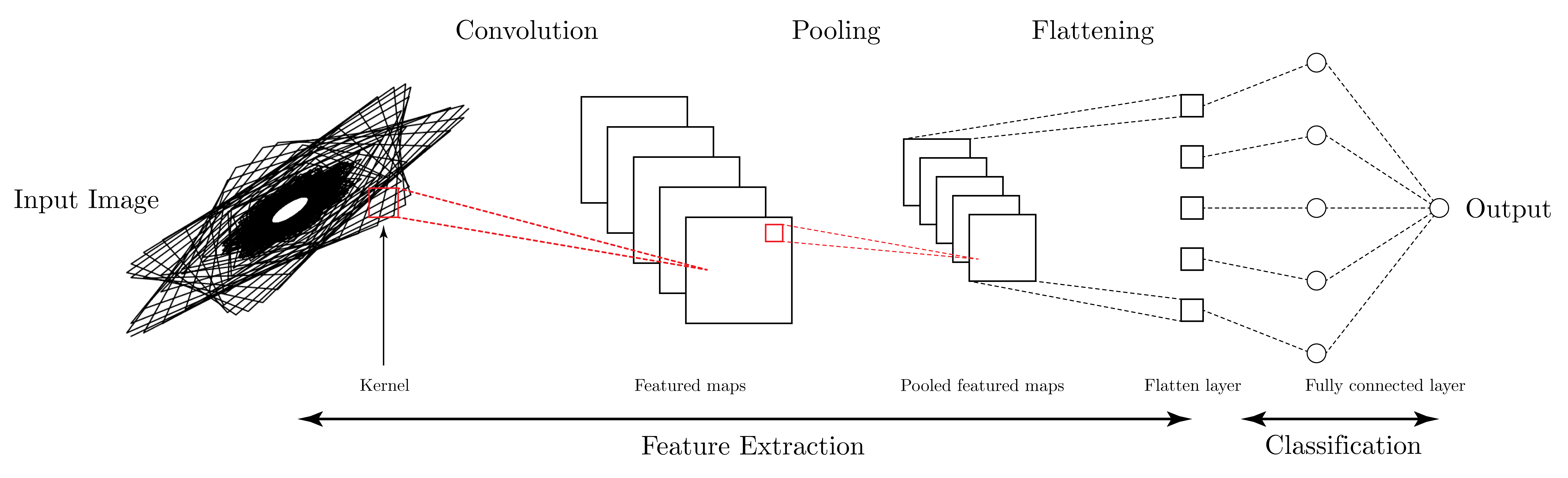}
    \caption{Convolution-based feature extraction and binary classification pipeline for phase space EEG images with no hidden layers in the classifier stage}
    \label{fig:cnn_arch}
\end{figure}

The generated phase space plots are subsequently processed through a Machine Learning Classifier, which is specifically designed to extract intricate spatial textures and temporal dynamics present in the EEG-derived images, as illustrated in Figure~\ref{fig:cnn_arch}. The model architecture is composed of multiple essential components, broadly categorized into two sections: feature extraction and classification. In our setup, only the feature extraction module of the model is retained, while the classification layer is replaced with external supervised learning classifiers for flexibility and better performance comparison.

In the feature extraction phase, each input phase space image, which is originally in RGB, is converted to grayscale to reduce dimensionality and translated into a NumPy array. A Laplacian filter of size $5 \times 5$ is then applied to the image to enhance edge-like structures and localized transitions that are particularly useful for differentiating between healthy and epileptic EEG patterns. The specific kernel used is defined as:

\[
\text{Laplace kernel}_{5\times5} = 
\begin{bmatrix}
  0  &  0  & -1  &  0  &  0 \\
  0  & -1  & -2  & -1  &  0 \\
 -1  & -2  &  16  & -2  & -1 \\
  0  & -1  & -2  & -1  &  0 \\
  0  &  0  & -1  &  0  &  0
\end{bmatrix}
\]

The convolution process involves sliding this kernel across the image data matrix using a defined stride, computing dot products between the kernel and the image patch at each position. The resulting values are aggregated into a feature map. To retain edge context at image boundaries, zero-padding is optionally added to the input image matrix.

Following convolution, a variable pooling strategy is applied to the feature maps using window sizes ranging from $2 \times 2$ to $16 \times 16$. This pooling operation, often using max-pooling, sub-samples the feature maps by extracting the maximum value within each window region, effectively reducing spatial dimensions and focusing on the most dominant activations. For example, a $16 \times 16$ max-pooling window with a stride value of 1 scans across the matrix to retain only the peak values within each block. This multi-scale pooling mechanism enables the model to detect features of varying sizes and resolutions across EEG representations. After pooling, the resultant feature maps are flattened into one-dimensional vectors (i.e., $1 \times 1200$), which serve as compact and discriminative feature representations. These flattened vectors are then passed to external classifiers for final decision-making.

To evaluate the discriminative strength of these extracted features, we independently experimented with two widely used classifiers: Logistic Regression and Support Vector Machine (SVM). Both models were trained separately on the same set of features to perform binary classification between healthy and epileptic EEG states. This comparative approach helps in identifying which classifier generalizes better under identical feature conditions.

\textbf{Logistic Regression} was configured with a sigmoid activation function to output probabilistic predictions. It computes a linear decision boundary by applying a weighted sum to the input features and adding a bias term. The resultant score is passed through the sigmoid function to yield a probability in the range $(0, 1)$:
\[
z = w^\top X + b, \quad \hat{y} = \frac{1}{1 + e^{-z}},
\]
where $X$ denotes the input feature vector, $w$ the weight vector, and $b$ the bias. The model is trained using the Binary Cross-Entropy (BCE) loss function:
\[
\mathcal{L}_{\text{BCE}} = -\frac{1}{N} \sum_{i=1}^{N} \left[y^{(i)} \log(\hat{y}^{(i)}) + (1 - y^{(i)}) \log(1 - \hat{y}^{(i)}) \right],
\]
which encourages the model to predict probabilities close to the true labels while penalizing misclassifications. Logistic Regression is particularly valued for its simplicity, interpretability, and ability to output calibrated confidence scores, making it suitable for threshold-based medical decision-making systems.

\textbf{Support Vector Machine (SVM)}, in contrast, was used as a non-probabilistic classifier to model complex, potentially nonlinear relationships in the high-dimensional feature space. A polynomial kernel was employed to implicitly map the feature space into a higher-dimensional domain, enabling the separation of nonlinearly separable data. The kernel function is defined as:
\[
K(X_i, X_j) = (\gamma X_i^\top X_j + r)^d,
\]
where $\gamma$ is a scaling parameter, $r$ is a kernel coefficient, and $d$ denotes the polynomial degree. The SVM classifier seeks the optimal hyperplane that maximizes the margin between the two classes:
\[
f(X) = \text{sign} \left( \sum_{i=1}^{N} \alpha_i y_i K(X_i, X) + b \right),
\]
where $\alpha_i$ represents the learned support vector coefficients, and $y_i$ and $X_i$ denote the training labels and feature vectors, respectively. While SVM does not provide probability estimates, it is highly effective in handling high-dimensional, imbalanced, and nonlinearly distributed data, which is often the case in EEG signal analysis.

The final classification output from each model is a binary label indicating whether the EEG input corresponds to a healthy or epileptic state. Notably, the overall architecture omits any deep fully connected layers after feature extraction. This design choice significantly reduces overfitting risks and ensures faster inference, which is particularly advantageous for resource-constrained or small-sample biomedical datasets.

This independent experimentation strategy—where Logistic Regression and SVM are trained and evaluated separately—provides a robust comparative analysis. It enables an informed evaluation of whether a linear probabilistic classifier like Logistic Regression or a margin-based kernel classifier like SVM is better suited to capture the distinctions embedded in EEG-derived phase space features. The effectiveness of both classifiers is further substantiated by qualitative analysis of the EEG power spectrum, where distinct gamma-band peaks (around 48–52 Hz and 74–78 Hz) were frequently observed in epileptic cases. These consistent spectral patterns reinforce the physiological relevance of the extracted features and form the basis for the statistical performance evaluation discussed in the following section.

\subsection{Evaluation Metrics for EEG Classification}
\begin{table}[ht!]
\centering
\renewcommand{\arraystretch}{1.5}
\caption{Summary of Evaluation Metrics}
\label{tab:metrics}
\begin{tabular}{|l|p{8.5cm}|c|}
\hline
\textbf{Metric} & \textbf{Description} & \textbf{Formula} \\
\hline
\textbf{Accuracy} & Proportion of all correctly classified instances among total samples. & $\frac{TP + TN}{TP + TN + FP + FN}$ \\
\hline
\textbf{Precision} & Proportion of predicted epileptic cases that are actually epileptic. & $\frac{TP}{TP + FP}$ \\
\hline
\textbf{Recall} & Ability to correctly identify actual epileptic cases. & $\frac{TP}{TP + FN}$ \\
\hline
\textbf{Specificity} & Ability to correctly identify actual normal (non-epileptic) cases. & $\frac{TN}{TN + FP}$ \\
\hline
\textbf{F1-Score} & Harmonic mean of precision and recall, balancing false positives and false negatives. & $2 \cdot \frac{\text{Precision} \cdot \text{Recall}}{\text{Precision} + \text{Recall}}$ \\
\hline
\end{tabular}
\end{table}
To assess the performance of the EEG classifier, key metrics are derived from the confusion matrix, defined as follows:
\begin{itemize}
    \item True Positive (TP): Epileptic events correctly classified as epileptic.
    \item False Positive (FP): Normal events incorrectly classified as epileptic.
    \item True Negative (TN): Normal events correctly classified as normal.
    \item False Negative (FN): Epileptic events incorrectly classified as normal.
\end{itemize}

Based on these terms, several standard evaluation metrics are computed as shown in Table \ref{tab:metrics}. These metrics together provide a holistic evaluation of the model's performance, particularly in medical diagnosis scenarios where both sensitivity and specificity are crucial.

\section{Results and Discussion}
\label{sec:result}
\subsection{Experimental setup}
We carried out this study on a system configured with an AMD Ryzen 5 5625U processor (6 cores, 12 threads), 16 GB of RAM, and integrated AMD $\text{Radeon}^{\text{TM}}$ graphics. The Miniconda 25.1.1 package was used to manage the computational environments based on Python 3.12 for dependencies such as Scikit-learn 1.6.1 for machine learning workflows, NumPy 2.2.3 for numerical computations, and Pandas 2.2.0 for data processing.

\subsection{Results}
\begin{figure}[H]
    \centering
    \includegraphics[width=0.48\linewidth]{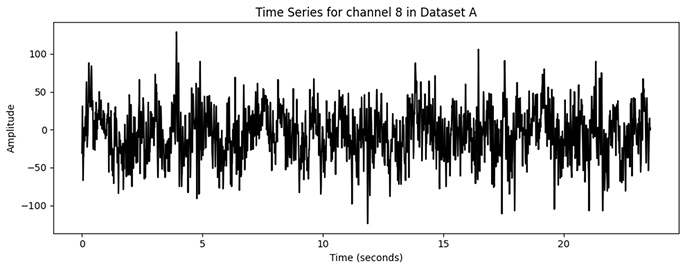}
    \hfill
    \includegraphics[width=0.48\linewidth]{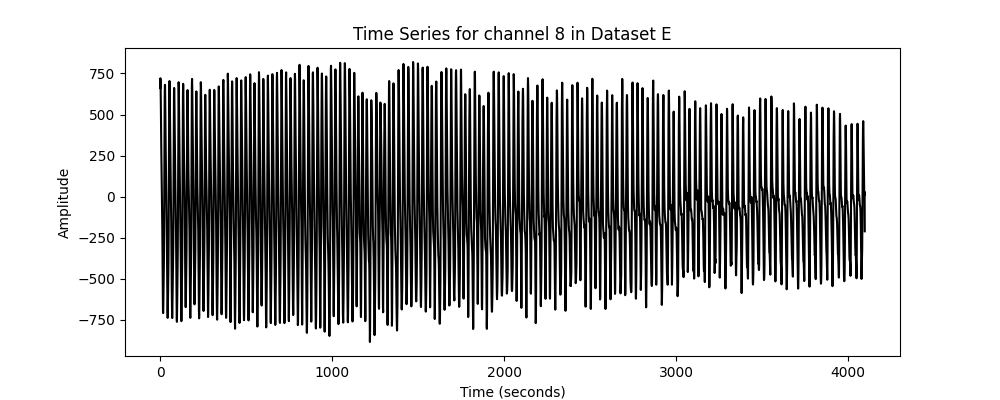}
    \caption{$8^{\text{th}}$ channel time series of EEG signal. \textbf{Left: }Dataset A (Healthy eyes open). \textbf{Right: } Dataset E (Epileptic ictal condition)}
    \label{fig:eeg_timeseries_comparison}
\end{figure}

The raw time series EEG signals from both healthy and epileptic subjects, as introduced in Section~\ref{subsec:dataset}, were first analyzed to understand baseline differences in neural dynamics. Each EEG recording comprises a single-channel signal of length $1 \times 4097$, sampled at 173.61~Hz, spanning approximately 23.6 seconds. Representative time series for the 8\textsuperscript{th} electrode channel from Dataset A (healthy, eyes open) and Dataset E (epileptic, ictal condition) are shown in Figure~\ref{fig:eeg_timeseries_comparison}. A clear distinction is evident in waveform morphology: while the healthy signal appears stochastic with low-amplitude oscillations, the ictal signal is dominated by high-amplitude rhythmic activity, characteristic of seizure onset.

% First image

\begin{figure}[h!]
    \centering
    \begin{minipage}[t]{0.5\textwidth}
        \centering
        \includegraphics[width=\linewidth]{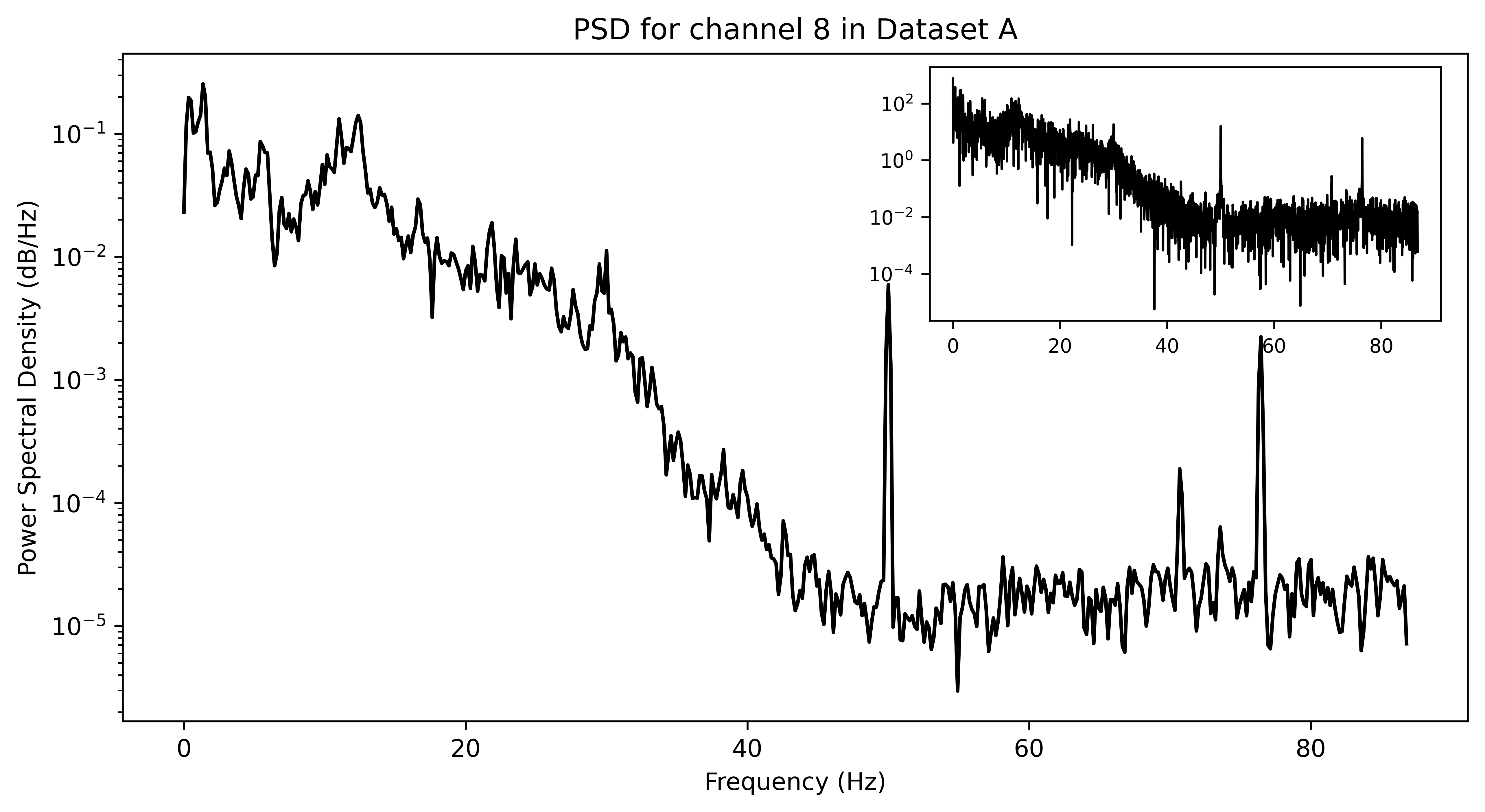}
    \end{minipage}%
    \begin{minipage}[t]{0.5\textwidth}
        \centering
        \includegraphics[width=\linewidth]{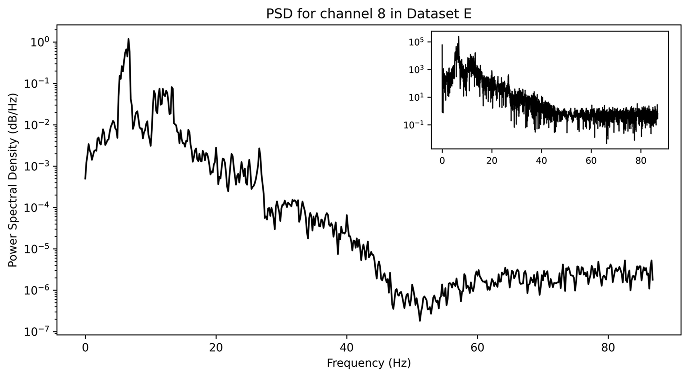}
    \end{minipage}
    \caption{Power spectral density analysis for 8\textsuperscript{th} channel using Welch transform. 
    \textbf{Left:} Healthy eyes open condition. \textbf{Right:} Seizure patient during ictal condition. 
    Both plots include \textbf{INSET} showing Fourier power spectrum.}
    \label{fig:psd_analysis}
\end{figure}

To probe underlying spectral characteristics, each signal was first standardized using z-score normalization and then subjected to spectral analysis. Two complementary methods were applied: the classical Fast Fourier Transform (FFT) for computing the raw power spectrum, and Welch’s method for smoothed Power Spectral Density (PSD) estimation. Welch's method employed a 50\% overlapping Hann window, averaging over 7 segments to reduce estimator variance, yielding a frequency resolution of approximately 0.17~Hz. The resulting PSD plots for the same channels in Datasets A and E are shown in Figure~\ref{fig:psd_analysis}, plotted on a semilogarithmic scale to highlight both high- and low-power frequency components. Inset panels within each plot also display the raw Fourier power spectrum for visual comparison.

\begin{figure}[h]
    \centering
    \includegraphics[width=0.5\linewidth]{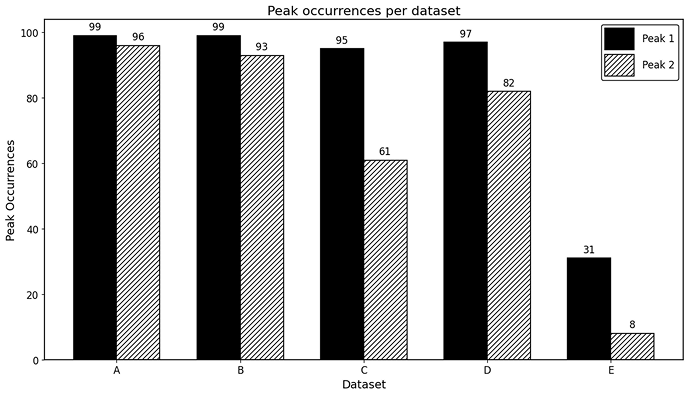}
    \caption{Statistics of channels with the first peak($48-52\text{Hz}$) and second peak($74-78\text{Hz}$) present in PSD plots of the datasets.}
    \label{fig:histogram_peaks}
\end{figure}

A notable finding from this spectral analysis was the differential presence of high-frequency gamma ($\gamma$) oscillations. While healthy EEG recordings (Dataset A) exhibited consistent peaks around the 48--52~Hz and 74--78~Hz ranges, these secondary gamma-band peaks were largely absent in ictal recordings (Dataset E). A statistical summary of peak occurrences across all datasets is provided in Figure~\ref{fig:histogram_peaks}, showing that nearly 99\% of channels in Dataset A exhibited both gamma-band peaks, whereas only 31\% and 8\% of Dataset E channels exhibited the first and second peaks, respectively. This suggests a significant attenuation of structured high-frequency components during seizure episodes, potentially indicating disrupted local circuit synchrony.

To investigate the dynamical behavior of EEG signals in specific gamma frequency bands, we performed a narrowband reconstruction of the signals using two spectral intervals identified during power spectral analysis: 48--52~Hz (first gamma peak) and 74--78~Hz (second gamma peak). These frequency bands were selected based on their consistent presence in healthy recordings and marked attenuation in ictal conditions, as revealed in the earlier PSD analysis.

The reconstruction was carried out by isolating the spectral components within each of these bands and applying an inverse Fast Fourier Transformation (IFFT). This allowed for the extraction of gamma-band-specific oscillatory behavior, enabling a focused analysis of frequency-dependent neural dynamics.

\begin{figure}[h!] % “Here, dammit!”
  \centering
  % Row 1
  \begin{subfigure}[b]{0.45\linewidth}
    \centering
    \includegraphics[width=\linewidth]{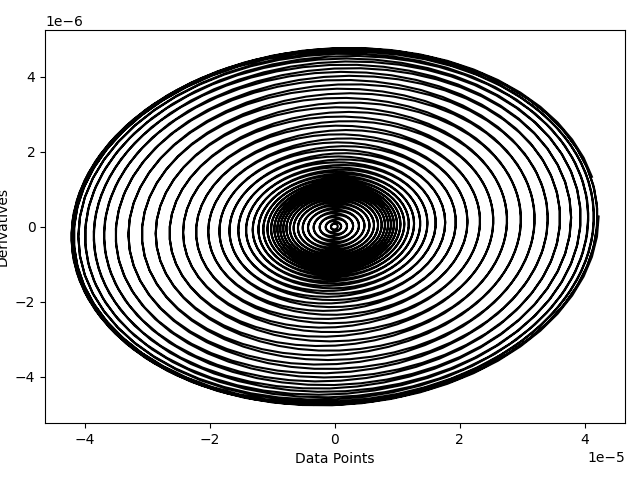}
    \caption{Healthy Eyes Open Phase Space}
  \end{subfigure}
  \hfill
  \begin{subfigure}[b]{0.45\linewidth}
    \centering
    \includegraphics[width=\linewidth]{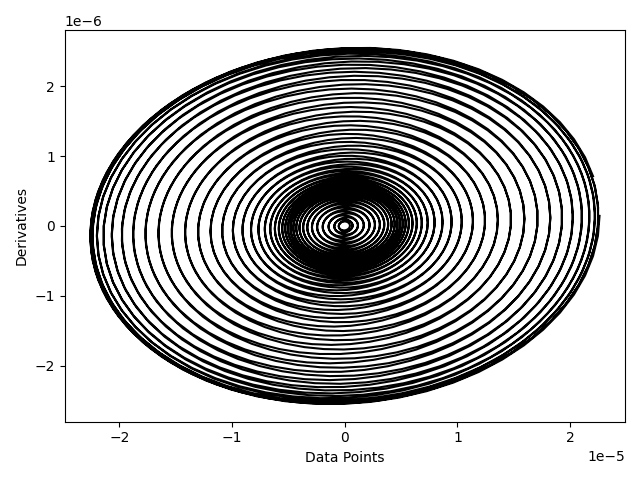}
    \caption{Healthy Eyes Closed Phase Space}
  \end{subfigure}

  \vspace{1em} % small vertical gap

  % Row 2
  \begin{subfigure}[b]{0.45\linewidth}
    \centering
    \includegraphics[width=\linewidth]{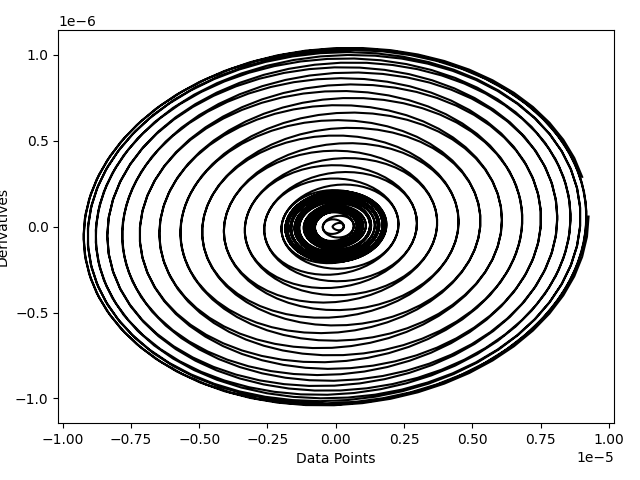}
    \caption{Epileptic Patient: Hippocampal Region}
  \end{subfigure}
  \hfill
  \begin{subfigure}[b]{0.45\linewidth}
    \centering
    \includegraphics[width=\linewidth]{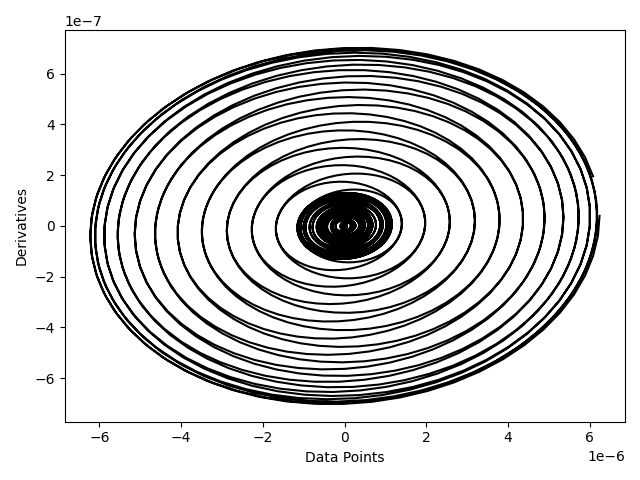}
    \caption{Epileptic Patient: Epileptogenic Region}
  \end{subfigure}

  \vspace{1em}

  % Row 3 (single, centered)
  \begin{subfigure}[b]{0.45\linewidth}
    \centering
    \includegraphics[width=\linewidth]{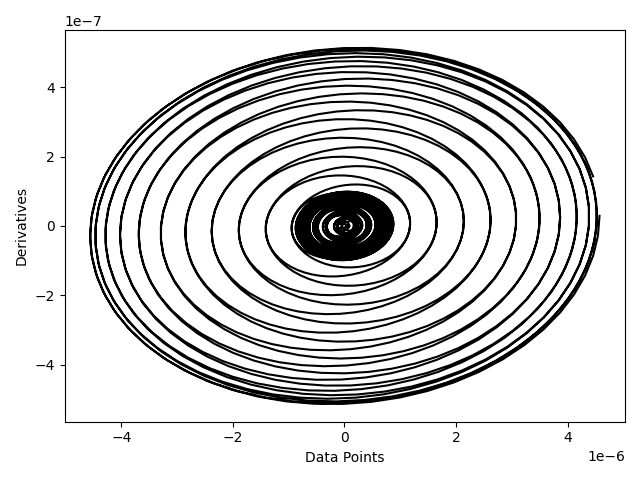}
    \caption{Epileptic Patient: Seizure Condition}
  \end{subfigure}

  \caption{Phase‐space analysis of the reconstructed signal from the first peak range (48–52\, Hz).}
  \label{fig:phase_space_open}
\end{figure}

%Second phase space analysis for the higher frequency range
\begin{figure}[h!] % Force placement HERE
    \centering
    % First row
    \begin{subfigure}[t]{0.45\textwidth}
        \includegraphics[width=\linewidth]{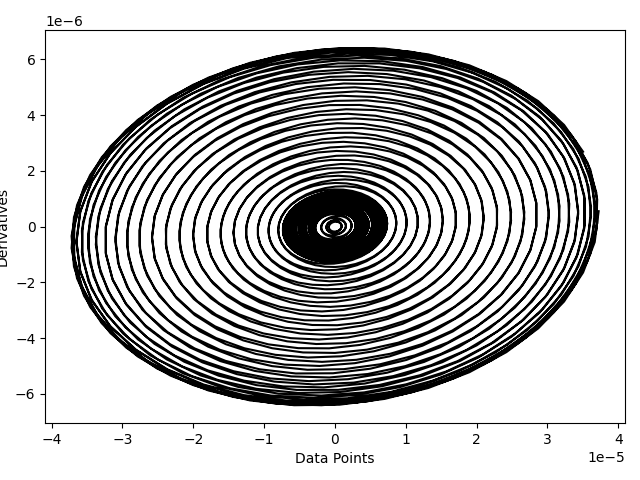}
        \caption{Healthy Eyes Open Phase Space}
        \label{fig:healthy_open}
    \end{subfigure}
    \hfill
    \begin{subfigure}[t]{0.45\textwidth}
        \includegraphics[width=\linewidth]{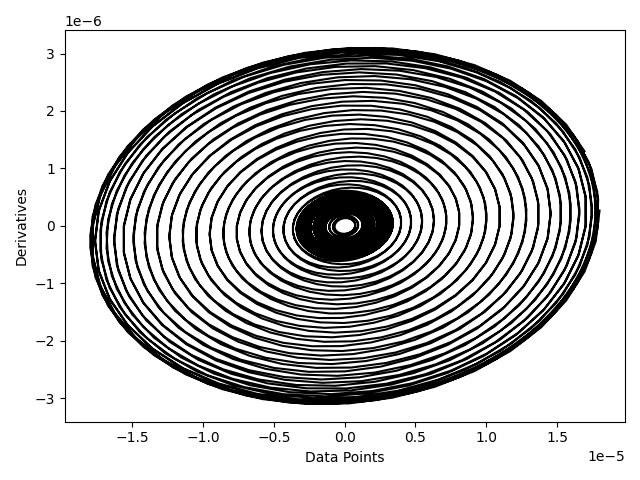}
        \caption{Healthy Eyes Closed Phase Space}
        \label{fig:healthy_closed}
    \end{subfigure}
    
    % Vertical space between rows
    \vspace{0.5cm}
    
    % Second row
    \begin{subfigure}[t]{0.45\textwidth}
        \includegraphics[width=\linewidth]{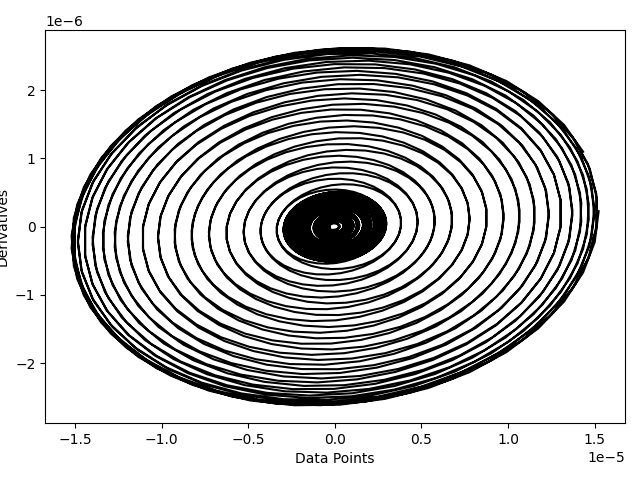}
        \caption{Epileptic patient (Hippocampal region)}
        \label{fig:hippocampal}
    \end{subfigure}
    \hfill
    \begin{subfigure}[t]{0.45\textwidth}
        \includegraphics[width=\linewidth]{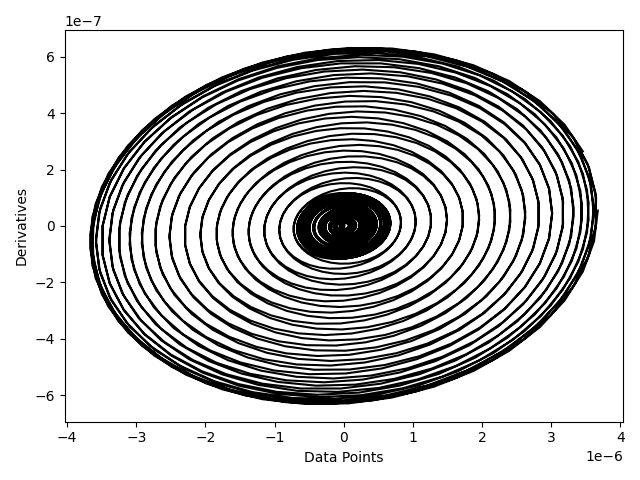}
        \caption{Epileptic patient (Epileptogenic region)}
        \label{fig:epileptogenic}
    \end{subfigure}
    
    % Vertical space between rows
    \vspace{0.5cm}
    
    % Third row (single-centered subfigure)
    \begin{subfigure}[t]{0.45\textwidth}
        \includegraphics[width=\linewidth]{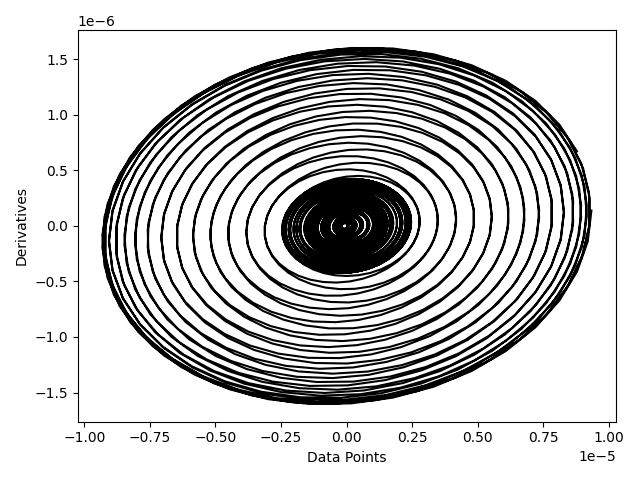}
        \caption{Epileptic patient (Ictal condition)}
        \label{fig:ictal}
    \end{subfigure}
    
    \caption{Phase space analysis of reconstructed signals from the second peak range (74-78 Hz)}
    \label{fig:phase_space_analysis}
\end{figure}

To visualize the temporal evolution of these reconstructed signals, we employed phase space analysis, where the amplitude of the signal is plotted against its first temporal derivative. This transformation yields two-dimensional trajectories that capture the system’s underlying geometry, revealing periodicity, bi-stability, or chaotic transitions in neural activity.

The phase space trajectories reconstructed from the 48--52~Hz band are shown in Figure~\ref{fig:phase_space_open}. Healthy subjects (Figures~\ref{fig:phase_space_open}a and \ref{fig:phase_space_open}b) exhibit structured, symmetric, and bi-stable orbits, suggesting stable rhythmic oscillations. Interictal epileptic states from the hippocampal and epileptogenic regions (Figures~\ref{fig:phase_space_open}c and \ref{fig:phase_space_open}d) demonstrate a partial loss of bi-stability, while the seizure condition (Figure~\ref{fig:phase_space_open}e) is characterized by irregular and merged orbits, indicative of bifurcation and nonlinear instability in cortical dynamics.

A similar analysis was conducted for the 74--78~Hz band, and the corresponding phase portraits are presented in Figure~\ref{fig:phase_space_analysis}. In this frequency range, healthy EEG segments again reveal regular cyclic orbits (Figures~\ref{fig:phase_space_analysis}a and \ref{fig:phase_space_analysis}b), while interictal signals show a reduced spread (Figures~\ref{fig:phase_space_analysis}c and \ref{fig:phase_space_analysis}d), suggesting decreased dynamical complexity. The ictal condition (Figure~\ref{fig:phase_space_analysis}e) is again marked by highly concentrated, collapsed orbits near the origin, reflecting strong attenuation and suppression of gamma oscillatory activity during seizures.

These results highlight the transition from well-structured, energy distributed oscillations in healthy states to spatially compressed and chaotic dynamics during seizure onset. To exploit these distinctive dynamical signatures, each phase portrait was converted into a 512 × 512 grayscale image and processed through an image classifier pipeline, as described in Section~\ref{Classification}.

The proposed classifier pipeline acted as a feature extractor, transforming the visual phase space patterns derived from gamma-band reconstructed signals into compact latent representations. These features were then used to train two traditional classifiers: Logistic Regression and Support Vector Machine (SVM). This hybrid approach was designed to assess whether simple classifiers, when provided with informative dynamical features, could reliably discriminate between physiological and pathological brain states.
%Hyperparameters used table
\begin{table}[h!]
\begin{center}
\begin{tabular}{|l|c|c|}
\hline
\textbf{Hyperparameter} & \textbf{Logistic Regression} & \textbf{SVM Model} \\
\hline
Optimizer            & Gradient Descent         & LibSVM (SMO) \\
Learning Rate        & 0.01                    & N/A \\
Batch Size           & Full batch (all data)    & N/A \\
Loss Function        & Binary Cross-Entropy     & Hinge Loss \\
Epochs (Iterations)  & 1000                    & N/A \\
Early Stopping       & Not used                & Not used \\
Kernel               & N/A                     & Polynomial (degree=3) \\
Regularization (C)   & N/A                     & Default (C=1.0) \\
\hline
\end{tabular}
\caption{Summary of Hyperparameters for Linear Regression and SVM Models}
\label{hyperparameters}
\end{center}
\end{table}

\begin{table}[h!]
    \centering
    % First row of tables (Linear Regression and SVM)

\begin{minipage}{\textwidth}
    \centering
    
    % Linear Regression Table
    \begin{tabular}{|c|c|c|c|c|c|}
        \hline
        \multicolumn{6}{|c|}{Linear Regression} \\
        \hline
        Evaluation Metrics & Results & AE & BE & CE & DE \\
        \hline
        \multirow{2}{*}{Accuracy} & Train & 0.93 & 0.85 & 0.75 & 0.89 \\ \cline{2-6}
         & Test & 0.85 & 0.72 & 0.75 & 0.80 \\
        \hline
        \multirow{2}{*}{Precision} & Train & 0.98 & 0.77 & 1 & 0.91  \\ \cline{2-6}
         & Test & 0.94 & 0.67 & 1 & 0.81 \\
        \hline
        \multirow{2}{*}{Recall} & Train &  0.88 & 0.98 & 0.5 & 0.85\\ \cline{2-6}
         & Test & 0.76 & 0.90 & 0.52 & 0.80 \\
        \hline
        \multirow{2}{*}{F1 Score} & Train & 0.93 & 0.87 & 0.66 & 0.78 \\ \cline{2-6}
         & Test & 0.84 & 0.77 & 0.68 & 0.82 \\
        \hline
        \multirow{2}{*}{Specificity} & Train & 0.98 & 0.72 & 1 & 0.92 \\ \cline{2-6}
         & Test & 0.94 & 0.52 & 1 & 0.79 \\
        \hline
    \end{tabular}
    
    \hspace{1em} % Horizontal spacing between tables
    
    % Support Vector Machine Table
    \begin{tabular}{|c|c|c|c|c|c|}
        \hline
        \multicolumn{6}{|c|}{Support Vector Machine} \\
        \hline
        Evaluation Metrics & Results & AE & BE & CE & DE \\
        \hline
        \multirow{2}{*}{Accuracy} & Train & 0.94 & 0.91 & 0.84 & 0.84 \\ \cline{2-6}
         & Test & 0.90 & 0.77 & 0.80 & 0.80 \\
        \hline
        \multirow{2}{*}{Precision} & Train & 0.92 & 0.87 & 0.90 & 0.93 \\ \cline{2-6}
         & Test & 0.90 & 0.73 & 0.88 & 0.88 \\
        \hline
        \multirow{2}{*}{Recall} & Train & 0.96 & 0.96 & 0.76 & 0.74 \\ \cline{2-6}
         & Test & 0.90 & 0.90 & 0.71 & 0.71 \\
        \hline
        \multirow{2}{*}{F1 Score} & Train & 0.94 & 0.91 & 0.83 & 0.82 \\ \cline{2-6}
         & Test & 0.90 & 0.80 & 0.78 & 0.78 \\
        \hline
        \multirow{2}{*}{Specificity} & Train & 0.92 & 0.86 & 0.92 & 0.95 \\ \cline{2-6}
         & Test & 0.89 & 0.63 & 0.89 & 0.89 \\
        \hline
    \end{tabular}
\end{minipage}
    % Vertical space between table rows
    \caption{\textbf{Model accuracy values for Peak 1 Welch (AE,BE,CE,DE)}. A: Healthy Eyes Open, B: Healthy Eyes Closed, C: Unhealthy Hippocampal region, D: Unhealthy Epileptogenic region, E: Unhealthy Ictal condition.}
    \label{tab:welch_peak1_type1}
\end{table}
The Logistic Regression classifier was optimized using gradient descent with a learning rate of 0.01, using binary cross-entropy loss and full-batch training over 1000 iterations without early stopping. The SVM model was implemented using LibSVM (SMO) with a polynomial kernel of degree 3 and default regularization (C = 1.0), and used hinge loss as its objective. A full summary of the hyperparameter configuration is presented in Table \ref{hyperparameters}.

Classification was carried out under both one-vs-rest and pairwise evaluation schemes using gamma peak frequency ranges of 48–52 Hz (Peak 1) and 74–78 Hz (Peak 2). The models were assessed using standard performance metrics, including accuracy, precision, recall, F1-score, and specificity, across both individual classes and class-pair combinations.

For the 48–52 Hz band (Peak 1), the results presented in Tables \ref{tab:welch_peak1_type1} and \ref{tab:welch_peak1_type2} reveal that the Support Vector Machine (SVM) consistently outperformed Logistic Regression in nearly all settings. In the one-vs-rest configuration (Table \ref{tab:welch_peak1_type1}), SVM achieved a test accuracy of 90\% and maintained F1-scores above 0.85 when classifying healthy (AE) versus ictal (DE) states. Logistic Regression also exhibited decent performance in specific classes such as CE and DE; however, it struggled with specificity, especially in distinguishing overlapping or ambiguous pathological conditions like BE.

The pairwise comparisons for Peak 1, detailed in Table \ref{tab:welch_peak1_type2}, further reinforce SVM’s superior classification ability. The model yielded higher precision and F1-scores across class pairs, particularly in separating pathological groups such as BC, BD, and CD. In contrast, Logistic Regression showed signs of overfitting to dominant classes, resulting in poor specificity (as low as 0.00 in several comparisons) and inflated recall scores that did not translate to reliable classification performance.

Moving to the 74–78 Hz range (Peak 2), the overall classification accuracy improved for both models, as evidenced in Tables \ref{tab:welch_peak2_type1} and \ref{tab:welch_peak2_type2}. In the one-vs-rest setup (Table \ref{tab:welch_peak2_type1}), SVM maintained robust generalization performance, delivering F1-scores above 0.80 in challenging pathological comparisons like BE vs. CE and CE vs. DE. Logistic Regression, while achieving higher accuracy in a few single-class evaluations, once again showed limitations in specificity, particularly in classes with complex temporal structure such as CE.

Pairwise classification results for Peak 2, summarized in Table \ref{tab:welch_peak2_type2}, confirmed the consistency and robustness of SVM. The model achieved balanced precision and recall values across nearly all class pairs, even in difficult comparisons involving overlapping epileptic states. Although Logistic Regression demonstrated some improvements in cases like AC and AD, its specificity values remained unstable, hindering reliable multi-class discrimination.

In summary, dynamical embeddings derived from gamma-band phase space portraits enabled classical machine learning models to effectively classify neural states. Among the classifiers tested, the Support Vector Machine (SVM) achieved the most favorable balance between sensitivity and specificity, demonstrating its superior suitability for detecting pathological state transitions within these nonlinear dynamics.

\begin{table}[h!]
    \centering
\begin{minipage}{\textwidth}
    \centering
    
% Linear Regression Table
\begin{tabular}{|c|c|c|c|c|c|c|c|}
    \hline
    \multicolumn{8}{|c|}{Linear Regression} \\
    \hline
    Evaluation Metrics & Results & AB & AC & AD & BC & BD & CD \\
    \hline
    \multirow{2}{*}{Accuracy} & Train   & 0.51 & 0.51 & 0.53 & 0.79 & 0.57 & 0.51 \\ \cline{2-8}
     & Test & 0.52 & 0.52 & 0.52 & 0.70 & 0.62 & 0.55 \\
    \hline
    \multirow{2}{*}{Precision} & Train & 0.50 & 0.50 & 0.51 & 0.80 & 0.53 & 0.50\\ \cline{2-8}
     & Test & 0.52 & 0.52 & 0.55 & 0.68 & 0.58 & 0.53 \\
    \hline
    \multirow{2}{*}{Recall} & Train & 1 & 1 & 1 & 0.78 & 1 & 1 \\ \cline{2-8}
     & Test & 1 & 1 & 1 & 0.80 & 1 & 1 \\
    \hline
    \multirow{2}{*}{F1 Score} & Train & 0.66 & 0.66 & 0.68 & 0.79 & 0.69 & 0.67 \\ \cline{2-8}
     & Test & 0.68 & 0.68 & 0.71 & 0.73 & 0.73 & 0.70 \\
    \hline
    \multirow{2}{*}{Specificity} & Train & 0.03 & 0.03 & 0.08 & 0.81 & 0.16 & 0.05 \\ \cline{2-8}
     & Test & 0 & 0 & 0.10 & 0.57 & 0.21 & 0.05 \\
    \hline
\end{tabular}

    \hspace{1em} % Horizontal spacing between tables
    
    % Support Vector Machine Table
\begin{tabular}{|c|c|c|c|c|c|c|c|}
    \hline
    \multicolumn{8}{|c|}{Support Vector Machine} \\
    \hline
    Evaluation Metrics & Results & AB & AC & AD & BC & BD & CD \\
    \hline
    \multirow{2}{*}{Accuracy} & Train   & 0.67 & 0.67 & 0.76 & 0.80 & 0.77 & 0.70 \\ \cline{2-8}
     & Test & 0.62 & 0.62 & 0.72 & 0.70 & 0.57 & 0.65 \\
    \hline
    \multirow{2}{*}{Precision} & Train & 0.77 & 0.77 & 0.75 & 0.78 & 0.75 & 0.69 \\ \cline{2-8}
     & Test & 0.71 & 0.71 & 0.75 & 0.66 & 0.57 & 0.64 \\
    \hline
    \multirow{2}{*}{Recall} & Train & 0.48 & 0.48 & 0.76 & 0.83 & 0.80 & 0.70 \\ \cline{2-8}
     & Test & 0.47 & 0.47 & 0.71 & 0.85 & 0.76 & 0.76 \\
    \hline
    \multirow{2}{*}{F1 Score} & Train & 0.59 & 0.59 & 0.76 & 0.80 & 0.78 & 0.70 \\ \cline{2-8}
     & Test & 0.57 & 0.57 & 0.73 & 0.75 & 0.65 & 0.69 \\
    \hline
    \multirow{2}{*}{Specificity} & Train &  0.86 & 0.86 & 0.76 & 0.77 & 0.75 & 0.70 \\ \cline{2-8}
     & Test & 0.78 & 0.78 & 0.73 & 0.52 & 0.36 & 0.52 \\
    \hline
\end{tabular}

\end{minipage}

    \caption{\textbf{Model accuracy values for Peak 1 Welch (AB, AC,AD,BC,BD,CD)}. A: Healthy Eyes Open, B: Healthy Eyes Closed, C: Unhealthy Hippocampal region, D: Unhealthy Epileptogenic region, E: Unhealthy Ictal condition.}
    \label{tab:welch_peak1_type2}
\end{table}

\begin{table}[h!]
    \centering
    % First row of tables (Linear Regression and SVM)
    \begin{minipage}{\textwidth}
        \centering

        % Linear Regression Table
        \begin{tabular}{|c|c|c|c|c|c|}
            \hline
            \multicolumn{6}{|c|}{Linear Regression} \\
            \hline
            Evaluation Metrics & Results & AE & BE & CE & DE \\
            \hline
            \multirow{2}{*}{Accuracy} &  Train   & 0.96 & 0.96 & 0.72 & 0.88\\ \cline{2-6}
             & Test  & 0.87 & 0.82 & 0.62 & 0.82  \\
            \hline
            \multirow{2}{*}{Precision} & Train & 0.97 & 0.97 & 0.64 & 0.87  \\ \cline{2-6}
             & Test & 0.86 & 0.85 & 0.58 & 0.85 \\
            \hline
            \multirow{2}{*}{Recall} & Train &  0.96 & 0.94 & 0.98 & 0.85 \\ \cline{2-6}
             & Test & 0.90 & 0.80 & 0.95 & 0.80 \\
            \hline
            \multirow{2}{*}{F1 Score} & Train & 0.96 & 0.96 & 0.78 & 0.88 \\ \cline{2-6}
             & Test & 0.88 & 0.82 & 0.72 & 0.82 \\
            \hline
            \multirow{2}{*}{Specificity} & Train & 0.97 & 0.97 & 0.47 & 0.87 \\ \cline{2-6}
             & Test & 0.84 & 0.84 & 0.26 & 0.84 \\
            \hline
        \end{tabular}

        \hspace{1em} % Horizontal spacing between tables

        % Support Vector Machine Table
        \begin{tabular}{|c|c|c|c|c|c|}
            \hline
            \multicolumn{6}{|c|}{Support Vector Machine} \\
            \hline
            Evaluation Metrics & Results & AE & BE & CE & DE \\
            \hline
            \multirow{2}{*}{Accuracy} & Train   & 0.93 & 0.93 & 0.80 & 0.86 \\ \cline{2-6}
             &  Test    & 0.85 & 0.82 & 0.77 & 0.80 \\
            \hline
            \multirow{2}{*}{Precision} & Train & 0.95 & 0.93 & 0.82 & 0.84 \\ \cline{2-6}
             & Test & 0.85 & 0.81 & 0.83 & 0.78 \\
            \hline
            \multirow{2}{*}{Recall} & Train & 0.89 & 0.92 & 0.82 & 0.88 \\ \cline{2-6}
             & Test & 0.85 & 0.85 & 0.71 & 0.85 \\
            \hline
            \multirow{2}{*}{F1 Score} & Train & 0.92 & 0.92 & 0.82 & 0.86 \\ \cline{2-6}
             & Test & 0.85 & 0.83 & 0.76 & 0.81 \\
            \hline
            \multirow{2}{*}{Specificity} & Train & 0.96 & 0.93 & 0.82 & 0.83 \\ \cline{2-6}
             & Test & 0.84 & 0.78 & 0.84 & 0.73 \\
            \hline
        \end{tabular}
    \end{minipage}

    \caption{\textbf{Model accuracy values for Peak 2 Welch (AE,BE,CE,DE)}. A: Healthy Eyes Open, B: Healthy Eyes Closed, C: Unhealthy Hippocampal region, D: Unhealthy Epileptogenic region, E: Unhealthy Ictal condition.}
    \label{tab:welch_peak2_type1}
\end{table}

\begin{table}[h!]
    \centering
    \begin{minipage}{\textwidth}
        \centering

        % Linear Regression Table
        \begin{tabular}{|c|c|c|c|c|c|c|c|}
            \hline
            \multicolumn{8}{|c|}{Linear Regression} \\
            \hline
            Evaluation Metrics & Results & AB & AC & AD & BC & BD & CD \\
            \hline
            \multirow{2}{*}{Accuracy} & Train   & 0.52 & 0.67 & 0.71 & 0.82 & 0.82 & 0.63\\ \cline{2-8}
             & Test    & 0.50 & 0.57 & 0.60 & 0.67 & 0.67 & 0.60 \\
            \hline
            \multirow{2}{*}{Precision} & Train & 1 & 0.60 & 0.65 & 0.78 & 0.78 & 0.57 \\ \cline{2-8}
             & Test & 1 & 0.55 & 0.58 & 0.63 & 0.65 & 0.57 \\
            \hline
            \multirow{2}{*}{Recall} & Train & 0.03 & 1 & 0.89 & 0.89 & 0.89 & 0.98 \\ \cline{2-8}
             & Test & 0.04 & 0.90 & 0.80 & 0.80 & 0.80 & 0.95 \\
            \hline
            \multirow{2}{*}{F1 Score} & Train & 0.07 & 0.75 & 0.75 & 0.83 & 0.83 & 0.72 \\ \cline{2-8}
             & Test & 0.09 & 0.69 & 0.68 & 0.72 & 0.72 & 0.71 \\
            \hline
            \multirow{2}{*}{Specificity} & Train & 1 & 0.35 & 0.53 & 0.76 & 0.76 & 0.30 \\ \cline{2-8}
             & Test & 1 & 0.21 & 0.36 & 0.52 & 0.52 & 0.21 \\
            \hline
        \end{tabular}

        \hspace{1em} % Horizontal spacing between tables

        % Support Vector Machine Table
        \begin{tabular}{|c|c|c|c|c|c|c|c|}
            \hline
            \multicolumn{8}{|c|}{Support Vector Machine} \\
            \hline
            Evaluation Metrics & Results & AB & AC & AD & BC & BD & CD \\
            \hline
            \multirow{2}{*}{Accuracy} & Train   & 0.72 & 0.82 & 0.72 & 0.83 & 0.83 & 0.77 \\ \cline{2-8}
             & Test    & 0.77 & 0.72 & 0.65 & 0.70 & 0.70 & 0.62\\
            \hline
            \multirow{2}{*}{Precision} & Train & 0.73 & 0.83 & 0.68 & 0.81 & 0.81 & 0.75 \\ \cline{2-8}
             & Test & 0.77 & 0.72 & 0.65 & 0.68 & 0.68 & 0.63 \\
            \hline
            \multirow{2}{*}{Recall} & Train & 0.67 & 0.79 & 0.79 & 0.86 & 0.86 & 0.80 \\ \cline{2-8}
             & Test & 0.80 & 0.76 & 0.71 & 0.80 & 0.80 & 0.66 \\
            \hline
            \multirow{2}{*}{F1 Score} & Train & 0.70 & 0.81 & 0.73 & 0.83 & 0.83 & 0.77 \\ \cline{2-8}
             & Test & 0.79 & 0.74 & 0.68 & 0.74 & 0.74 & 0.65 \\
            \hline
            \multirow{2}{*}{Specificity} & Train & 0.76 & 0.85 & 0.65 & 0.81 & 0.81 & 0.73 \\ \cline{2-8}
             & Test & 0.73 & 0.68 & 0.57 & 0.57 & 0.58 & 0.58 \\
            \hline
        \end{tabular}
    \end{minipage}

    \caption{\textbf{Model accuracy values for Peak 2 Welch (AB,AC,AD,BC,BD,CD)}. A: Healthy Eyes Open, B: Healthy Eyes Closed, C: Unhealthy Hippocampal region, D: Unhealthy Epileptogenic region, E: Unhealthy Ictal condition.}
    \label{tab:welch_peak2_type2}
\end{table}

\subsection{Comparison with State-of-art methods}
Table~\ref{tab:research_papers_1} and \ref{tab:research_papers} presents a comparative analysis of recent seizure classification approaches on the UoB dataset. While most prior works rely on complex deep architectures such as CNN-LSTM, ResBiLSTM, or BiLSTM-GRU models, they typically operate on raw EEG signals or standard time–frequency representations like STFT and DWT \cite{9754583, 10.3389/fnins.2024.1436619, ILIAS2023119010, 10.3389/fncom.2024.1415967, 10.3389/fncom.2024.1340251, 10.3389/fncom.2024.1415967}.
In contrast, the proposed method introduces a distinctive feature representation by generating phase-space images from EEG signals reconstructed using peak frequencies of the power spectral density (PSD). This approach captures the underlying nonlinear and dynamical behavior of epileptic brain states in a compact visual form. By transforming spectral dynamics into geometric patterns, the method offers an interpretable alternative to deep feature embeddings. Despite using classical classifiers such as SVM and logistic regression, the system achieves a competitive accuracy of 94\%, highlighting the effectiveness of this feature extraction in distinguishing seizure and non-seizure states. The simplicity, interpretability, and efficiency of the approach make it a strong candidate for real-time and low-resource deployment.

\begin{table}[H]
\caption{State-of-the-art seizure classification studies using UoB dataset}
\label{tab:research_papers_1}
\begin{tabular}{|>{\raggedright\arraybackslash}p{2.0cm}|>{\raggedright\arraybackslash}p{1.5cm}|>{\raggedright\arraybackslash}p{4.5cm}|>{\raggedright\arraybackslash}p{2.2cm}|>{\raggedright\arraybackslash}p{2.0cm}|}
\hline
\textbf{Author} & \textbf{Year} & \textbf{Methodology} & \textbf{Accuracy\%} & \textbf{Classifier} \\
\hline
Shankar et al \cite{SHANKAR2021102854} & Aug 2021 & Recurrence Plot image generation from EEG delta rhythm & 93 & CNN \\
\hline
Woodbright et al \cite{WOODBRIGHT202130} & July 2021 & Rule generation from extracted CNN deep features & 98.65 & Decision Trees \\
\hline
Duan et al \cite{Duan2022An} & Dec 2021 & Deep metric learning with 1D CNN feature extractor & 98.6 & Deep Metric Learning \\
\hline
Tuncer et al \cite{TUNCER2022103462} & March 2022 & Hyperparameter optimization of the Bi-LSTM model & 99 & Bi-LSTM \\
\hline
Xin et al \cite{9754583} & April 2022 & Multi-scale wavelet decomposition & 98.89 & Attention based Wavelet CNN \\
\hline
Shen et al \cite{SHEN2022103820} & Aug 2022 & DB4-DWT with eigenvalue based feature extraction & 97 & SVM \\
\hline
Tran et al \cite{diagnostics12112879} & Nov 2022 & DWT-based statistical feature extraction with BPSO for feature selection & 98.4 & SVM \\
\hline
Ilias et al \cite{ILIAS2023119010} & March 2023 & Multimodal CNN with a GMU to fuse features from raw and STFT based EEG & 98.75 & Gated Multimodal DNN \\
\hline
Liu et al \cite{LIU2023104693} & May 2023 & Semi-JMI feature selection & 97.5 & AFM-DQN \\
\hline
Yang et al \cite{10.3389/fncom.2023.1294770} & Nov 2023 & Transfer learning with an attention-based, temporal-spectral feature model & 97.5 & Attention based CNN \\
\hline
\end{tabular}
\end{table}

\begin{table}[h!]
\caption{State-of-the-art seizure classification studies using UoB dataset}
\label{tab:research_papers}
\begin{tabular}{|>{\raggedright\arraybackslash}p{2.0cm}|>{\raggedright\arraybackslash}p{1.5cm}|>{\raggedright\arraybackslash}p{4.8cm}|>{\raggedright\arraybackslash}p{2.2cm}|>{\raggedright\arraybackslash}p{2.0cm}|}
\hline
\textbf{Author} & \textbf{Year} & \textbf{Methodology} & \textbf{Accuracy\%} & \textbf{Classifier} \\
\hline
Skaria et al \cite{skaria} & Mar 2024 & Phase Space Reconstruction (PSR) with Elliptical Features & 100 & KNN \\
\hline
Mallick et al \cite{10.3389/fncom.2024.1340251} & Mar 2024 & 1D-CNN + BiLSTM/GRU (recurrent unit) & 95.81–100 & Hybrid CNN-RNN \\
\hline
Abdullayeva et al \cite{Abdullayeva2024} & April 2024 & Welch extraction method + RF\slash LSTM\slash SVM\slash NB\slash LM & 97.66–99.87 & RF\slash LSTM\slash SVM\slash NB\slash LM \\
\hline
Wang et al \cite{s24113360} & May 2024 & Multimodal Dual-Stream (1D/2D-CNN + LSTM + CA) & 99.69 & Hybrid Neural Network \\
\hline
Zhao et al \cite{10.3389/fncom.2024.1415967} & Jun 2024 & ResBiLSTM (1D-ResNet + BiLSTM) & 98.88–100 & ResBiLSTM \\
\hline
Nie et al \cite{10.3389/fnins.2024.1436619} & Jul 2024 & Fast Fourier Transform extraction + CNN-LSTM & 97.95–99.83 & CNN-LSTM \\
\hline
Elshekhidris et al \cite{elshekhidris} & Feb 2025 & 1D-CNN with EMD + SWT denoising & 100 & 1D-CNN \\
\hline
Berrich et al \cite{Berrich} & Apr 2025 & Hybrid CNN/DNN-SVM with PCA dimensionality reduction & 96.97–99.96 & SVM \\
\hline
Atlam et al \cite{app15094690} & Apr 2025 & SMOTE + PCA-DWT hybrid feature selection & 97.3 & SVM \\
\hline
Farawan et al \cite{Farawn} & Jun 2025 & Multiple feature selection (MDI/Corr/SFS/SBS) & 98.1–98.6 & Random Forest \\
\hline
Karthik et al \cite{article} & Jun 2025 & DWT-based feature extraction (time, frequency, entropy) & 97 & SVM \\
\hline
\textbf{This Work} & 2025 & Classify phase-space images from signal PSD peak frequencies & 94 & SVM, LR (sigmoid) \\
\hline
\end{tabular}
\end{table}

\section{Conclusion}
\label{sec:conclusion}
% We have identified the characteristic pattern of changes in the collective gamma wave modulations, occurring in epileptic patients as compared to normal ones, as well as the changes in their physical behavior during the onset of epilepsy. This is done through a combination of the Welch transform to isolate the inherently noisy high-frequency gamma wave modulations and employing a phase space approach for pinpointing the nature of changes in the diseased conditions. The pattern of changes is physically explainable, leading to the employment of image classification based on a CNN approach for differentiation.

We investigated the nature of the modifications in the temporal dynamics manifest in the high-frequency EEG spectra of the normal human brain in comparison to the diseased brain undergoing epilepsy. For this purpose, the Fourier reconstruction is efficaciously made use of after the Welch transform helped identify the relevant frequency components undergoing significant changes in the case of epilepsy.
The temporal dynamics involved in the EEG signals and their associated variations showed a well-structured periodic pattern characterised by bistability and significant quantifiable structural changes during epileptic episodes. A support vector machine is found to discern the differences and identify the onsets of epilepsy with up to 94 percent accuracy.
The partial reconstruction of the dynamics, as compared to the earlier studies of the full phase space, accurately pinpointed the destabilisation of the collective high-frequency synchronous behaviour and its precise differences in the normal and epileptic individuals.

%----suggested section

\subsection{Abbreviations}
DWT: Discrete Wavelet Transform; MDI: Mean Decrease Impurity; Corr: Correlation; SFS: Sequential Forward Selection; SBS: Sequential Backward Selection; CNN: Convolutional Neural Network; DNN: Deep Neural Network; SVM: Support Vector Machine; PCA: Principal Component Analysis; SMOTE: Synthetic Minority Over-sampling Technique; EMD: Empirical Mode Decomposition; SWT: Stationary Wavelet Transform; LSTM: Long Short-Term Memory; ResBiLSTM: Residual Bidirectional Long Short-Term Memory; 1D-ResNet: One-Dimensional Residual Network; BiLSTM: Bidirectional Long Short-Term Memory; NB: Naive Bayes; LM: Linear Model; GRU: Gated Recurrent Unit; KNN: k-Nearest Neighbor

\subsection{Data Availability}
The EEG dataset used in this study is publicly available University of Bonn EEG dataset \cite{PhysRevE.64.061907}. The dataset can be accessed directly at \url{https://www.upf.edu/web/ntsa/downloads/-/asset_publisher/xvT6E4pczrBw/content/2001-indications-of-nonlinear-deterministic-and-finite-dimensional-structures-in-time-series-of-brain-electrical-activity-dependence-on-recording-regi}. It is provided exclusively for research and educational use.

\subsection{Author Contributions}
\textbf{Jyotiraj Nath:} Methodology, Software, Validation, Formal Analysis, Investigation, Data Curation, Writing – Original Draft, Visualization.\\\textbf{Shreya Banerjee:} Formal Analysis, Data Curation, Software, Investigation, Writing – Review \& Editing.\\\textbf{Bhaswati Singha Deo:} Writing – Original Draft, Writing – Review \& Editing.\\\textbf{Mayukha Pal:} Methodology, Writing - Review.\\\textbf{Prasanta K. Panigrahi:} Conceptualization, Supervision, Project Administration, Resources, Writing – Review \& Editing.

\subsection{Acknowledgments}
Mayukha Pal would like to thank ABB for their support in this work.

\subsection{Conflicts of Interest}
The authors have no conflicts of interest to disclose.
% \section{References}
\bibliographystyle{elsarticle_num} 
\bibliography{main}

\begin{thebibliography}{10}
\expandafter\ifx\csname url\endcsname\relax
  \def\url#1{\texttt{#1}}\fi
\expandafter\ifx\csname urlprefix\endcsname\relax\def\urlprefix{URL }\fi
\expandafter\ifx\csname href\endcsname\relax
  \def\href#1#2{#2} \def\path#1{#1}\fi

\bibitem{WHO2023}
W.~H. Organization, \href{https://www.who.int/news-room/fact-sheets/detail/epilepsy}{Epilepsy: A public health imperative}, WHO Report (2023).
\newline\urlprefix\url{https://www.who.int/news-room/fact-sheets/detail/epilepsy}

\bibitem{MALAKOUTI2025134}
S.~M. Malakouti, \href{https://www.sciencedirect.com/science/article/pii/S258891412500019X}{Enhanced epilepsy detection using discrete wavelet transform and bandpass filtering on eeg data: integration of art-based and lvq models}, Clinical eHealth 8 (2025) 134--145.
\newblock \href {https://doi.org/https://doi.org/10.1016/j.ceh.2025.07.001} {\path{doi:https://doi.org/10.1016/j.ceh.2025.07.001}}.
\newline\urlprefix\url{https://www.sciencedirect.com/science/article/pii/S258891412500019X}

\bibitem{Schomer2018}
D.~L. Schomer, F.~H. Lopes~da Silva (Eds.), Niedermeyer's Electroencephalography: Basic Principles, Clinical Applications, and Related Fields, seventh Edition, Oxford University Press, 2018.

\bibitem{Moezzi2022}
B.~Moezzi, S.~Chiang, J.~Gotman, Gamma oscillations and interictal epileptiform discharges: A review of recent findings, Clinical Neurophysiology 137 (2022) 76--87.
\newblock \href {https://doi.org/10.1016/j.clinph.2022.01.007} {\path{doi:10.1016/j.clinph.2022.01.007}}.

\bibitem{Tsiouris2021}
K.~Tsiouris, V.~Pezoulas, M.~Zervakis, A long short-term memory deep learning network for the prediction of epileptic seizures using eeg signals, Computers in Biology and Medicine 134 (2021) 104793.
\newblock \href {https://doi.org/10.1016/j.compbiomed.2021.104793} {\path{doi:10.1016/j.compbiomed.2021.104793}}.

\bibitem{9714339}
M.~Pal, Neeraj, P.~K. Panigrahi, Coupled oscillator dynamics in brain eeg signals: Manifestation of synchronization and across frequency energy exchange by neutral turbulence, IEEE Access 10 (2022) 20445--20454.
\newblock \href {https://doi.org/10.1109/ACCESS.2022.3151692} {\path{doi:10.1109/ACCESS.2022.3151692}}.

\bibitem{9754583}
Q.~Xin, S.~Hu, S.~Liu, L.~Zhao, Y.-D. Zhang, An attention-based wavelet convolution neural network for epilepsy eeg classification, IEEE Transactions on Neural Systems and Rehabilitation Engineering 30 (2022) 957--966.
\newblock \href {https://doi.org/10.1109/TNSRE.2022.3166181} {\path{doi:10.1109/TNSRE.2022.3166181}}.

\bibitem{SHEN2022103820}
M.~Shen, P.~Wen, B.~Song, Y.~Li, \href{https://www.sciencedirect.com/science/article/pii/S1746809422003421}{An eeg based real-time epilepsy seizure detection approach using discrete wavelet transform and machine learning methods}, Biomedical Signal Processing and Control 77 (2022) 103820.
\newblock \href {https://doi.org/https://doi.org/10.1016/j.bspc.2022.103820} {\path{doi:https://doi.org/10.1016/j.bspc.2022.103820}}.
\newline\urlprefix\url{https://www.sciencedirect.com/science/article/pii/S1746809422003421}

\bibitem{10.3389/fnins.2024.1436619}
J.~Nie, H.~Shu, F.~Wu, \href{https://www.frontiersin.org/journals/neuroscience/articles/10.3389/fnins.2024.1436619}{An epilepsy classification based on fft and fully convolutional neural network nested lstm}, Frontiers in Neuroscience Volume 18 - 2024 (2024).
\newblock \href {https://doi.org/10.3389/fnins.2024.1436619} {\path{doi:10.3389/fnins.2024.1436619}}.
\newline\urlprefix\url{https://www.frontiersin.org/journals/neuroscience/articles/10.3389/fnins.2024.1436619}

\bibitem{PAL2022126516}
M.~Pal, M.~P., P.~K. Panigrahi, \href{https://www.sciencedirect.com/science/article/pii/S0378437121007895}{A multi scale time–frequency analysis on electroencephalogram signals}, Physica A: Statistical Mechanics and its Applications 586 (2022) 126516.
\newblock \href {https://doi.org/https://doi.org/10.1016/j.physa.2021.126516} {\path{doi:https://doi.org/10.1016/j.physa.2021.126516}}.
\newline\urlprefix\url{https://www.sciencedirect.com/science/article/pii/S0378437121007895}

\bibitem{kannathal2005entropies}
N.~Kannathal, M.~L. Choo, U.~R. Acharya, P.~Sadasivan, \href{https://www.sciencedirect.com/science/article/pii/S0169260705001525}{Entropies for detection of epilepsy in eeg}, Computer Methods and Programs in Biomedicine 80~(3) (2005) 187--194.
\newblock \href {https://doi.org/https://doi.org/10.1016/j.cmpb.2005.06.012} {\path{doi:https://doi.org/10.1016/j.cmpb.2005.06.012}}.
\newline\urlprefix\url{https://www.sciencedirect.com/science/article/pii/S0169260705001525}

\bibitem{rosso2001wavelet}
O.~A. Rosso, S.~Blanco, J.~Yordanova, V.~Kolev, A.~Figliola, M.~Schürmann, E.~Başar, \href{https://www.sciencedirect.com/science/article/pii/S0165027000003563}{Wavelet entropy: a new tool for analysis of short duration brain electrical signals}, Journal of Neuroscience Methods 105~(1) (2001) 65--75.
\newblock \href {https://doi.org/https://doi.org/10.1016/S0165-0270(00)00356-3} {\path{doi:https://doi.org/10.1016/S0165-0270(00)00356-3}}.
\newline\urlprefix\url{https://www.sciencedirect.com/science/article/pii/S0165027000003563}

\bibitem{Farawn}
A.~Al~Farawn, S.~Al-Khammasi, A.~Alhilali, N.~Ali, Eeg feature selection techniques for epileptic seizure detection: Performance and evaluation study, International Journal of Mathematics, Statistics, and Computer Science 3 (2025) 345--358.
\newblock \href {https://doi.org/10.59543/ijmscs.v3i.14771} {\path{doi:10.59543/ijmscs.v3i.14771}}.

\bibitem{Duan2022An}
L.~Duan, Z.~Wang, Y.~Qiao, Y.~Wang, Z.~Huang, B.~Zhang, An automatic method for epileptic seizure detection based on deep metric learning, IEEE Journal of Biomedical and Health Informatics 26~(5) (2022) 2147--2157.
\newblock \href {https://doi.org/10.1109/JBHI.2021.3138852} {\path{doi:10.1109/JBHI.2021.3138852}}.

\bibitem{elshekhidris}
I.~Elshekhidris, M.~Amien, F.~Ahmed, Automated classification of epileptic seizures using modified one-dimensional convolution neural network based on empirical mode decomposition with high accuracy, Neuroscience Informatics 5 (2025) 100188.
\newblock \href {https://doi.org/10.1016/j.neuri.2025.100188} {\path{doi:10.1016/j.neuri.2025.100188}}.

\bibitem{s24113360}
B.~Wang, Y.~Xu, S.~Peng, H.~Wang, F.~Li, \href{https://www.mdpi.com/1424-8220/24/11/3360}{Detection method of epileptic seizures using a neural network model based on multimodal dual-stream networks}, Sensors 24~(11) (2024).
\newblock \href {https://doi.org/10.3390/s24113360} {\path{doi:10.3390/s24113360}}.
\newline\urlprefix\url{https://www.mdpi.com/1424-8220/24/11/3360}

\bibitem{10541111}
X.~Zhou, C.~Liu, R.~Yang, L.~Zhang, L.~Zhai, Z.~Jia, Y.~Liu, Learning robust global-local representation from eeg for neural epilepsy detection, IEEE Transactions on Artificial Intelligence 5~(11) (2024) 5720--5732.
\newblock \href {https://doi.org/10.1109/TAI.2024.3406289} {\path{doi:10.1109/TAI.2024.3406289}}.

\bibitem{Jia2020}
X.~Jia, L.~Wang, Power spectral density and neural dynamics in epilepsy: Challenges and perspectives, Journal of Neuroscience Methods 338 (2020) 108682.
\newblock \href {https://doi.org/10.1016/j.jneumeth.2020.108682} {\path{doi:10.1016/j.jneumeth.2020.108682}}.

\bibitem{Moca2021}
V.~Moca, A.~Nagy-Darabont, R.~Muresan, Time-frequency representations for the analysis of neural oscillations: a practical guide, Journal of Neuroscience Methods 347 (2021) 108951.
\newblock \href {https://doi.org/10.1016/j.jneumeth.2020.108951} {\path{doi:10.1016/j.jneumeth.2020.108951}}.

\bibitem{Li2021}
Y.~Li, W.~Zhang, Z.~Zhang, Phase-space analysis of eeg signals for seizure detection and prediction, Biomedical Signal Processing and Control 69 (2021) 102903.
\newblock \href {https://doi.org/10.1016/j.bspc.2021.102903} {\path{doi:10.1016/j.bspc.2021.102903}}.

\bibitem{Roy2022}
S.~Roy, M.~K. Islam, A.~Sadiq, Machine learning and deep learning approaches for epileptic seizure detection using eeg: A review, Sensors 22~(2) (2022) 472.
\newblock \href {https://doi.org/10.3390/s22020472} {\path{doi:10.3390/s22020472}}.

\bibitem{Ahmad2023}
I.~Ahmad, M.~Raza, M.~J. Khan, A hybrid cnn–lstm model for seizure detection from eeg signals, Biomedical Signal Processing and Control 80 (2023) 104313.
\newblock \href {https://doi.org/10.1016/j.bspc.2023.104313} {\path{doi:10.1016/j.bspc.2023.104313}}.

\bibitem{PhysRevE.64.061907}
R.~G. Andrzejak, K.~Lehnertz, F.~Mormann, C.~Rieke, P.~David, C.~E. Elger, \href{https://link.aps.org/doi/10.1103/PhysRevE.64.061907}{Indications of nonlinear deterministic and finite-dimensional structures in time series of brain electrical activity: Dependence on recording region and brain state}, Phys. Rev. E 64 (2001) 061907.
\newblock \href {https://doi.org/10.1103/PhysRevE.64.061907} {\path{doi:10.1103/PhysRevE.64.061907}}.
\newline\urlprefix\url{https://link.aps.org/doi/10.1103/PhysRevE.64.061907}

\bibitem{article_jasper}
H.~Jasper, The ten-twenty electrode system of the international federation of electroencephalography, Electronceph Clin Neurophysiol 10 (1958) 371--375.

\bibitem{doi:https://doi.org/10.1002/9780470511923.ch2}
S.~Sanei, J.~Chambers, \href{https://onlinelibrary.wiley.com/doi/abs/10.1002/9780470511923.ch2}{Fundamentals of EEG Signal Processing}, John Wiley \& Sons, Ltd, 2007, Ch.~2, pp. 35--125.
\newblock \href {http://arxiv.org/abs/https://onlinelibrary.wiley.com/doi/pdf/10.1002/9780470511923.ch2} {\path{arXiv:https://onlinelibrary.wiley.com/doi/pdf/10.1002/9780470511923.ch2}}, \href {https://doi.org/https://doi.org/10.1002/9780470511923.ch2} {\path{doi:https://doi.org/10.1002/9780470511923.ch2}}.
\newline\urlprefix\url{https://onlinelibrary.wiley.com/doi/abs/10.1002/9780470511923.ch2}

\bibitem{https://doi.org/10.1111/j.0013-9580.2005.66104.x}
R.~S. Fisher, W.~v.~E. Boas, W.~Blume, C.~Elger, P.~Genton, P.~Lee, J.~Engel~Jr., \href{https://onlinelibrary.wiley.com/doi/abs/10.1111/j.0013-9580.2005.66104.x}{Epileptic seizures and epilepsy: Definitions proposed by the international league against epilepsy (ilae) and the international bureau for epilepsy (ibe)}, Epilepsia 46~(4) (2005) 470--472.
\newblock \href {http://arxiv.org/abs/https://onlinelibrary.wiley.com/doi/pdf/10.1111/j.0013-9580.2005.66104.x} {\path{arXiv:https://onlinelibrary.wiley.com/doi/pdf/10.1111/j.0013-9580.2005.66104.x}}, \href {https://doi.org/https://doi.org/10.1111/j.0013-9580.2005.66104.x} {\path{doi:https://doi.org/10.1111/j.0013-9580.2005.66104.x}}.
\newline\urlprefix\url{https://onlinelibrary.wiley.com/doi/abs/10.1111/j.0013-9580.2005.66104.x}

\bibitem{widmann2015digital}
A.~Widmann, E.~Schröger, B.~Maess, \href{https://www.sciencedirect.com/science/article/pii/S0165027014002866}{Digital filter design for electrophysiological data – a practical approach}, Journal of Neuroscience Methods 250 (2015) 34--46, cutting-edge EEG Methods.
\newblock \href {https://doi.org/https://doi.org/10.1016/j.jneumeth.2014.08.002} {\path{doi:https://doi.org/10.1016/j.jneumeth.2014.08.002}}.
\newline\urlprefix\url{https://www.sciencedirect.com/science/article/pii/S0165027014002866}

\bibitem{oppenheim1999discrete}
A.~V. Oppenheim, R.~W. Schafer, J.~R. Buck, Discrete-time signal processing, Prentice hall, 1999.

\bibitem{welch1967use}
P.~Welch, The use of fast fourier transform for the estimation of power spectra: A method based on time averaging over short, modified periodograms, IEEE Transactions on Audio and Electroacoustics 15~(2) (1967) 70--73.
\newblock \href {https://doi.org/10.1109/TAU.1967.1161901} {\path{doi:10.1109/TAU.1967.1161901}}.

\bibitem{tenke2015digital}
J.~Kayser, C.~E. Tenke, \href{https://www.sciencedirect.com/science/article/pii/S0167876015001609}{Issues and considerations for using the scalp surface laplacian in eeg/erp research: A tutorial review}, International Journal of Psychophysiology 97~(3) (2015) 189--209, on the benefits of using surface Laplacian (current source density) methodology in electrophysiology.
\newblock \href {https://doi.org/https://doi.org/10.1016/j.ijpsycho.2015.04.012} {\path{doi:https://doi.org/10.1016/j.ijpsycho.2015.04.012}}.
\newline\urlprefix\url{https://www.sciencedirect.com/science/article/pii/S0167876015001609}

\bibitem{stam2005nonlinear}
C.~Stam, \href{https://www.sciencedirect.com/science/article/pii/S1388245705002403}{Nonlinear dynamical analysis of eeg and meg: Review of an emerging field}, Clinical Neurophysiology 116~(10) (2005) 2266--2301.
\newblock \href {https://doi.org/https://doi.org/10.1016/j.clinph.2005.06.011} {\path{doi:https://doi.org/10.1016/j.clinph.2005.06.011}}.
\newline\urlprefix\url{https://www.sciencedirect.com/science/article/pii/S1388245705002403}

\bibitem{takens1981detecting}
F.~Takens, Detecting strange attractors in turbulence, in: D.~Rand, L.-S. Young (Eds.), Dynamical Systems and Turbulence, Warwick 1980, Springer Berlin Heidelberg, Berlin, Heidelberg, 1981, pp. 366--381.

\bibitem{kantz2004nonlinear}
H.~Kantz, T.~Schreiber, Nonlinear time series analysis, Cambridge university press, 2004.

\bibitem{ILIAS2023119010}
L.~Ilias, D.~Askounis, J.~Psarras, \href{https://www.sciencedirect.com/science/article/pii/S0957417422020280}{Multimodal detection of epilepsy with deep neural networks}, Expert Systems with Applications 213 (2023) 119010.
\newblock \href {https://doi.org/https://doi.org/10.1016/j.eswa.2022.119010} {\path{doi:https://doi.org/10.1016/j.eswa.2022.119010}}.
\newline\urlprefix\url{https://www.sciencedirect.com/science/article/pii/S0957417422020280}

\bibitem{10.3389/fncom.2024.1415967}
W.~Zhao, W.-F. Wang, L.~M. Patnaik, B.-C. Zhang, S.-J. Weng, S.-X. Xiao, D.-Z. Wei, H.-F. Zhou, \href{https://www.frontiersin.org/journals/computational-neuroscience/articles/10.3389/fncom.2024.1415967}{Residual and bidirectional lstm for epileptic seizure detection}, Frontiers in Computational Neuroscience Volume 18 - 2024 (2024).
\newblock \href {https://doi.org/10.3389/fncom.2024.1415967} {\path{doi:10.3389/fncom.2024.1415967}}.
\newline\urlprefix\url{https://www.frontiersin.org/journals/computational-neuroscience/articles/10.3389/fncom.2024.1415967}

\bibitem{10.3389/fncom.2024.1340251}
S.~Mallick, V.~Baths, \href{https://www.frontiersin.org/journals/computational-neuroscience/articles/10.3389/fncom.2024.1340251}{Novel deep learning framework for detection of epileptic seizures using eeg signals}, Frontiers in Computational Neuroscience Volume 18 - 2024 (2024).
\newblock \href {https://doi.org/10.3389/fncom.2024.1340251} {\path{doi:10.3389/fncom.2024.1340251}}.
\newline\urlprefix\url{https://www.frontiersin.org/journals/computational-neuroscience/articles/10.3389/fncom.2024.1340251}

\bibitem{SHANKAR2021102854}
A.~Shankar, H.~K. Khaing, S.~Dandapat, S.~Barma, \href{https://www.sciencedirect.com/science/article/pii/S1746809421004511}{Analysis of epileptic seizures based on eeg using recurrence plot images and deep learning}, Biomedical Signal Processing and Control 69 (2021) 102854.
\newblock \href {https://doi.org/https://doi.org/10.1016/j.bspc.2021.102854} {\path{doi:https://doi.org/10.1016/j.bspc.2021.102854}}.
\newline\urlprefix\url{https://www.sciencedirect.com/science/article/pii/S1746809421004511}

\bibitem{WOODBRIGHT202130}
M.~Woodbright, B.~Verma, A.~Haidar, \href{https://www.sciencedirect.com/science/article/pii/S0925231221003088}{Autonomous deep feature extraction based method for epileptic eeg brain seizure classification}, Neurocomputing 444 (2021) 30--37.
\newblock \href {https://doi.org/https://doi.org/10.1016/j.neucom.2021.02.052} {\path{doi:https://doi.org/10.1016/j.neucom.2021.02.052}}.
\newline\urlprefix\url{https://www.sciencedirect.com/science/article/pii/S0925231221003088}

\bibitem{TUNCER2022103462}
E.~Tuncer, E.~{Doğru Bolat}, \href{https://www.sciencedirect.com/science/article/pii/S1746809421010594}{Classification of epileptic seizures from electroencephalogram (eeg) data using bidirectional short-term memory (bi-lstm) network architecture}, Biomedical Signal Processing and Control 73 (2022) 103462.
\newblock \href {https://doi.org/https://doi.org/10.1016/j.bspc.2021.103462} {\path{doi:https://doi.org/10.1016/j.bspc.2021.103462}}.
\newline\urlprefix\url{https://www.sciencedirect.com/science/article/pii/S1746809421010594}

\bibitem{diagnostics12112879}
L.~V. Tran, H.~M. Tran, T.~M. Le, T.~T.~M. Huynh, H.~T. Tran, S.~V.~T. Dao, \href{https://www.mdpi.com/2075-4418/12/11/2879}{Application of machine learning in epileptic seizure detection}, Diagnostics 12~(11) (2022).
\newblock \href {https://doi.org/10.3390/diagnostics12112879} {\path{doi:10.3390/diagnostics12112879}}.
\newline\urlprefix\url{https://www.mdpi.com/2075-4418/12/11/2879}

\bibitem{LIU2023104693}
X.~Liu, X.~Ding, J.~Liu, W.~Nie, Q.~Yuan, \href{https://www.sciencedirect.com/science/article/pii/S174680942300126X}{Automatic focal eeg identification based on deep reinforcement learning}, Biomedical Signal Processing and Control 83 (2023) 104693.
\newblock \href {https://doi.org/https://doi.org/10.1016/j.bspc.2023.104693} {\path{doi:https://doi.org/10.1016/j.bspc.2023.104693}}.
\newline\urlprefix\url{https://www.sciencedirect.com/science/article/pii/S174680942300126X}

\bibitem{10.3389/fncom.2023.1294770}
Y.~Yang, F.~Li, J.~Luo, X.~Qin, D.~Huang, \href{https://www.frontiersin.org/journals/computational-neuroscience/articles/10.3389/fncom.2023.1294770}{Epileptic focus localization using transfer learning on multi-modal eeg}, Frontiers in Computational Neuroscience Volume 17 - 2023 (2023).
\newblock \href {https://doi.org/10.3389/fncom.2023.1294770} {\path{doi:10.3389/fncom.2023.1294770}}.
\newline\urlprefix\url{https://www.frontiersin.org/journals/computational-neuroscience/articles/10.3389/fncom.2023.1294770}

\bibitem{skaria}
S.~Skaria, S.~Savithryamma, Automatic classification of seizure and seizure-free eeg signals based on phase space reconstruction features, Journal of Biological Physics 50 (2024) 1--16.
\newblock \href {https://doi.org/10.1007/s10867-024-09654-6} {\path{doi:10.1007/s10867-024-09654-6}}.

\bibitem{Abdullayeva2024}
E.~Abdullayeva, H.~K. Örnek, \href{https://doi.org/10.18280/ts.410237}{Diagnosing epilepsy from {EEG} using machine learning and {Welch} spectral analysis}, Traitement du Signal 41~(2) (2024) 971--977.
\newblock \href {https://doi.org/10.18280/ts.410237} {\path{doi:10.18280/ts.410237}}.
\newline\urlprefix\url{https://doi.org/10.18280/ts.410237}

\bibitem{Berrich}
Y.~Berrich, Z.~Guennoun, Eeg-based epilepsy detection using cnn-svm and dnn-svm with feature dimensionality reduction by pca, Scientific Reports 15 (04 2025).
\newblock \href {https://doi.org/10.1038/s41598-025-95831-z} {\path{doi:10.1038/s41598-025-95831-z}}.

\bibitem{app15094690}
H.~F. Atlam, G.~E. Aderibigbe, M.~S. Nadeem, \href{https://www.mdpi.com/2076-3417/15/9/4690}{Effective epileptic seizure detection with hybrid feature selection and smote-based data balancing using svm classifier}, Applied Sciences 15~(9) (2025).
\newblock \href {https://doi.org/10.3390/app15094690} {\path{doi:10.3390/app15094690}}.
\newline\urlprefix\url{https://www.mdpi.com/2076-3417/15/9/4690}

\bibitem{article}
S.~Karthik, K.~Bharath, B.~Ramji, K.~Puttegowda, A.~Gowda, S.~D~S, Enhanced eeg signal processing for accurate epileptic seizure detection, SN Computer Science 6 (07 2025).
\newblock \href {https://doi.org/10.1007/s42979-025-04148-1} {\path{doi:10.1007/s42979-025-04148-1}}.

\end{thebibliography}

\end{document}